\documentclass[preprint] {imsart}
\RequirePackage[OT1]{fontenc}
\RequirePackage{amsthm, amsmath, amssymb, amsfonts, epsfig, subfigure, multirow,color}
\RequirePackage[authoryear]{natbib}
\RequirePackage[colorlinks=true,citecolor=blue,urlcolor=blue]{hyperref}

\startlocaldefs
\numberwithin{equation}{section}
\theoremstyle{plain}
\def\M{\mathbb M}

\def\R{\mathbb R}

\def\G{\mathbb G}
\def\P{\mathbb P}

\def\be{\begin{equation}}
\def\ee{\end{equation}}
\def\qed{\hfill$\square$}
\def\ed{\end{document}}

\newtheorem{prop}{Proposition}
\newtheorem{thm}{Theorem}
\newtheorem{cor}{Corollary}
\newtheorem{lem}{Lemma}
\newtheorem{rmk}{Remark}

\endlocaldefs

\begin{document}
\sloppy
\begin{frontmatter}
\title{Baseline zone estimation in two dimensions} 
\runtitle{Baseline zone estimation}

\begin{aug}
\author{\fnms{Atul} \snm{Mallik}\thanksref{t1}\ead[label=e1]{atulm@umich.edu}},
\author{\fnms{Moulinath} \snm{Banerjee}\thanksref{t1}\ead[label=e2]{moulib@umich.edu}}
\and
\author{\fnms{Michael} \snm{Woodroofe}\ead[label=e3]{michaelw@umich.edu}}

\thankstext{t1}{Supported by NSF Grant DMS-1007751  and a Sokol Faculty Award, University 
of Michigan}
\runauthor{A. Mallik, M. Banerjee and M. Woodroofe}

\address{Department of Statistics\\
University of Michigan\\
Ann Arbor, Michigan 48109 \\
\phantom{E-mail:\ }\printead*{e1}
\phantom{E-mail:\ }\printead*{e2}
\phantom{E-mail:\ }\printead*{e3}}

\end{aug}

\begin{abstract}We consider the problem of estimating the region on which a non-parametric regression function is at its baseline level in two dimensions. The baseline level typically corresponds to the minimum/maximum of the function and estimating such regions or their complements is pertinent to several problems arising in edge estimation, environmental statistics, fMRI and related fields. We assume the baseline region to be convex and estimate it via fitting a ``stump'' function to approximate $p$-values obtained from tests for deviation of the regression function from its baseline level. The estimates, obtained using an algorithm originally developed for constructing convex contours of a density, are studied in two different sampling settings, one where several responses can be obtained at a number of different covariate-levels (dose--response)
and the other involving limited number of response values per covariate (standard regression). The shape of the baseline region and the smoothness of the regression function at its boundary play a critical role in determining the rate of convergence of our estimate: for a regression function which is ``$p$-regular'' at the boundary of the convex baseline region, our estimate converges  at a rate $N^{-2/(4p+3)}$ in the dose--response setting, $N$ being the total budget, and its analogue in the standard regression setting converges at a rate of $N^{-1/(2p +2)}$. Extensions to non-convex baseline regions are explored as well.
\end{abstract}



\end{frontmatter}

\section{Introduction}
Consider a data generating model of the form
$ Y = \mu (X) + \epsilon,$
where $\mu$ is a function on $[0,1]^2$ such that 
\be
\label{eq:mubase}
\mu(x) = \tau_0 \mbox{ for } x \in S_0, \mbox{ and } \mu (x) > \tau_0 \mbox{ for } x \notin S_0 
\ee 
and $\tau_0$ is unknown. The covariate $X$ may arise from a random or a fixed design setting and we assume that $\epsilon$ has mean zero with finite positive variance $\sigma^2_0$. We are interested in estimating the baseline region $S_0$ beyond which the function deviates from its baseline value. There are several practical motivations behind detecting $S_0$ (or $S_0^c$) which can be thought of as the region of no-signal. For example, in several {\it fMRI} studies, one seeks to detect regions of brain activity from cross sectional two-dimensional images. Here, $S_0$ corresponds to the region of no-activity in the brain with $S_0^c$ being the region of interest. In {\it LIDAR (light detection and ranging)} experiments used for measuring concentration of pollutants in the atmosphere, interest often centers on finding high/low pollution zones (see, for example, \cite{WM86}); in such contexts, $S_0$ would be the zone of minimal pollution. In {\it dose-response studies}, patients may be put on multiple (interacting) drugs (see, for example, \cite{GB09}), and it is of interest to find the dosage levels ($\partial S_0$) at which the effect of the drugs starts kicking in. 

The question of detecting $S_0$ is also related to the \emph{edge detection} problem which involves recovering the boundary of an image. In edge detection, $\mu$ corresponds to the image intensity function with $S_0^c$ being the image and $S_0$ the background. A number of different algorithms in the computer science literature deal with this problem, though primarily in situations where $\mu$ has a jump discontinuity at the boundary of $S_0$; see \cite{Qu07} for a review of edge detection techniques. With the exception of work done by \cite{KsT93}, \cite{MmT95} and a few others, theoretical properties of such algorithms appear to have been rarely addressed. In fact, the study of theoretical properties of such estimates is typically intractable without some regularity assumption on $S_0$; for example, \cite{MmT95} discuss minimax recovery of sets under smoothness assumptions on the boundary. 

In this paper, we approach the problem from the point of view of a shape-constraint (typically obtained from background knowledge) on the baseline region. We assume that the region $S_0$ is a closed convex subset of $[0,1]^2$ with a non-empty interior (and therefore, positive Lebesgue measure) and restrict ourselves to the more difficult problem where $\mu$ is continuous at the boundary. Convexity is a natural shape restriction to impose, not only because of analytical tractability, but also as convex boundaries arise naturally in several application areas: see, \cite{convex-wang}, \cite{convex-guo}, \cite{convex-stahl} and \cite{convex-golden} for a few illustrative examples. In the statistics literature, \cite{GldnZ06} provide theoretical analyses of a convex boundary recovery method in a white noise framework. While this has natural connections to our problem, we note that they impose certain conditions (see Definitions 2 and 3 of \cite{GldnZ06} and the associated discussions), which restricts the geometry of the set of interest, $G$, beyond convexity. Hence, their results, particularly on the rate of convergence, are difficult to compare to the ones obtained in our problem. Further, they estimate $G$ through its \emph{support function} which needs to be estimated along all directions. It is unclear whether an effective algorithm can be devised to adopt this procedure in a regression setting. 

Our problem also has connections to the level-sets estimation problem since $S_0^c$ is the ``level-set'' $\{ x: \mu(x) > \tau_0 \}$ of the function $\mu$. However, because $\tau_0$ is at the extremity of the range of $\mu$, the typical level-set estimate $\{ x: \hat{\mu} (x) > \tau_0 \}$, where $\hat{\mu}$ is an estimate of $\mu$, does not perform well unless $\mu$ has a jump at $\partial S_0$ (a situation not considered in this paper). Moreover, this plug-in approach does not account for the pre-specified shape of the level-set. 
 We note that the shape-constrained approach to estimate level-sets has received some attention in the literature, e.g., \citet{N91} studied estimating ellipsoidal level-sets in the context of densities, \citet{H87} provided an algorithm for estimating convex contours of a density, and \cite{T97} and \cite{C97}  studied ``star-shaped'' level-sets of density and regression functions respectively. All the above approaches are based on an ``excess mass'' criterion (or its local version) that yield estimates with  optimal convergence rates \citep{T97}. It will be seen later that our estimate also recovers the level-set of a transform of $\mu$, but at a level in the \emph{interior} of the range of the transform. More connections in this regard are explored in Section \ref{sec:lvl}. 

In this paper, we extend the approach of \cite{ABMG10}, developed in a simple 1-dimensional setting, to obtain an estimate of $S_0$. We construct $p$-value type statistics which detect the deviation of the function $\mu$ from its baseline value $\tau_0$ at each covariate level and then fit an appropriate ``stump'' -- a piecewise constant function with two levels -- to these $p$-values. We study the problem in two distinct sampling settings: the so called `dose--response' setting where plenty of replicates are available at each covariate value, and the (standard) regression setting where limited (taken to be 1 without loss of generality) responses are available at each covariate level. As mentioned earlier, the regression setting is pertinent to several compelling applications. The dose-response setting is motivated by the minimum effective dose (MED) problems (a one-dimensional version of our problem) where data are available from several patients (multiple replicates) at each dose level (covariate value) and one is interested in finding the lowest dose level where the effect of the concerned drug kicks in. The baseline set in this case is, therefore, an interval $[0, d_0]$ for some unknown $d_0 >0$.  The extension 
of the dose-response setting to two dimensions not only provides theoretical insight into the behavior of our procedure but is also relevant to pharmacological studies involving drugs that interact. 

The smoothness of $\mu$ at its boundary plays a critical role in determining the rate of convergence of our estimate: for a regression function which is ``$p$-regular'' (formally defined in Section \ref{sec:note}) at the boundary of the convex baseline region, our estimate converges  at a rate $N^{-2/(4p+3)}$ (Theorem \ref{c5prp:ratemu} and the following remark) in the dose--response setting, $N$ being the total budget. This coincides with the minimax rate of a related level-set estimation problem; see  \cite[Theorem 3.7]{P95} and \cite[Theorem 2]{T97}. The analogue of the estimate in the regression setting converges at the slightly slower rate of $N^{-1/(2p +2)}$ (Theorem \ref{c5prp:ratemureg}). The difference in the two rates is due to the bias introduced from the use of kernel estimates in the regression setting. A more technical explanation is given in Remark \ref{rm:bias}. It should be pointed out that our convergence rates are very different from the analogous problem in the density estimation scenario which corresponds to finding the support of a multivariate density. Faster convergence rates \citep{HPT95}can be obtained in density estimation due to the simpler nature of the problem: namely, there are \emph{no realizations} from outside the support of the density.

The main contributions of the paper are the following. We propose a novel and computationally simple approach to estimate baseline sets in two dimensions and deduce consistency and rates of convergence of our estimate in the two aforementioned settings. Our approach falls at the interface of edge detection and level-set estimation problems as it detects the edge set ($S_0^c$) through a level-set estimate (see Section \ref{sec:lvl}). The proofs require heavy-duty applications of non-standard empirical processes  and, along the way, we deduce results which may be of independent interest. For example, we apply a blocking argument which leads to a version of Hoeffding's inequality for $m$-dependent random fields, which is then further extended to an empirical process inequality. This should find usage in spatial statistics and is potentially relevant to approaches based on $m$-approximations  that answer the central limit question for dependent random fields and their empirical process extensions; see \cite{Rsn69}, \cite{Bol82} and \cite{YW13} for some work on $m$-dependent random fields and $m$-approximations.  While we primarily address the situation where the baseline set is convex, in the presence of efficient algorithms, our approach is extendible beyond convexity (see Section \ref{sec:lvl}). 
 
The rest of the paper is arranged as follows: we formally define the two settings  and describe the estimation procedure in Section \ref{desc}. Barring $\mu$ and $S_0$, notations are \emph{not carried forward} from the dose-response setting to the regression setting unless stated otherwise. We list our assumptions in Section \ref{sec:note}. We justify consistency and deduce an upper bound on the rate of the convergence of our procedure (assuming a known $\tau_0$) for the dose-response and regression settings in Sections \ref{dsrsp2} and \ref{regrs2} respectively.  Situations with unknown $\tau_0$ are addressed in Section \ref{sec:extntau2}. We explore extensions to non-convex baseline regions and connections with level-set estimation in Section \ref{sec:lvl}.

\section{Estimation Procedure}\label{desc}
In this section, we develop a multi-dimensional version of a $p$-value procedure originally developed in a one-dimensional setting in \cite{ABMG10}. 
\subsection{Dose-Response Setting}\label{doseset}
Consider a data generating model of the form 
\begin{eqnarray}\label{eq:dsr}
Y_{i j} = \mu(X_{i}) + \epsilon_{ij}, \ \ j = 1, \ldots, m, \ \ i = 1,\ldots,n.\nonumber
\end{eqnarray}
Here $m = m_n = m_0 n^\beta$ for some $\beta >0$, with $N = m \times n$ being the total budget.
The covariate $X$ is sampled from a distribution $F$ with  Lebesgue density $f$ on $[0,1]^2$ and $\epsilon$ is independent of $X$, has mean 0 and variance $\sigma_0^2$. 

At each level $X_i = x$, we test the null hypothesis $H_{0,x}: \mu(x) = \tau_0$ against the alternative $H_{1,x}: \mu(x) > \tau_0$ and use the resulting (approximate) $p$-values to construct an estimate of the set $S_0$. The \emph{non-normalized} $p$-values are given by $$p_{m,n} (x) = 1 - \Phi(\sqrt{m}(\bar Y_{i\cdot} - \hat{\tau})),$$ where $\bar Y_{i\cdot} = \sum_{j=1}^m Y_{ij}/m$ and $\hat{\tau}$ is some suitable estimate of $\mu$ (to be discussed later). 
These $p$-values asymptotically have mean 1/2 for $x \in S_0$ and converge to zero when $x \notin S_0$. This simple observation can be used to construct estimates of $S_0$. We fit a stump to the observed $p$--values, with levels 1/2 and 0 on either side of the boundary of the set and prescribe the set corresponding to the best fitting stump (in the sense of least squares) as an estimate of $S_0$. Formally, we define $\xi_S(x) = (1/2) 1 (x \in S)$ and we  minimize
\begin{eqnarray}\label{eq:CPSSEv1}
\sum_{i=1}^n \left( p_{m,n} (X_i) - \xi_S(X_i)\right)^2= \sum_{i: X_i \in S} \left( p_{m,n}(X_i) - \frac{1}{2} \right)^2 +  \sum_{i: X_i \in S^c} \left(p_{m,n}(X_i)\right)^2 \nonumber
\end{eqnarray}
over choices of $S$.  
The above least squares problem can be reduced to minimizing 
\begin{eqnarray}\label{eq:CPSSE}
\M_n(S) = \P_n \left\{ \Phi\left(\sqrt{m}(\bar{Y} - \hat{\tau})\right) - \gamma\right\} 1_{S}(X), \nonumber
\end{eqnarray}
where $\P_n$ denotes the empirical measure on $\{ \bar Y_{i\cdot} , X_i \}_{i\leq n}$ and $\gamma = 3/4$. 
\begin{rmk}
\label{defend-yours} 
Our methodology uses non-normalized $p$-values, since the test statistic sitting inside the argument to $\Phi$ has not been normalized by the estimate of the variance. Alternatively, one could have considered fitting a stump to the normalized $p$-values. This alternative version of the procedure would exhibit the same fundamental feature, namely, dichotomous separation  over $S_0$ and $S_0^c$ which is why the stump-based procedure works, and produce identical rates of convergence. The non-normalized version is analytically and notationally more tractable as it avoids some routine (but tedious) algebraic justifications required for the normalized version.
\end{rmk} 
The class of sets over which  $\M_n$ is minimized  should be chosen carefully as very large classes would give uninteresting discrete sets while small classes may not provide a reasonable estimate of $S_0$. As we assumed $S_0$ to be convex, we minimize $\M_n$ over $\mathcal{S}$, the class of \emph{closed convex} subsets of $[0,1]^2$. Let $\hat{S}_n = \mathop{\mbox{argmin}}_{S \in \mathcal S} \M_n (S).$ The estimate $\hat{S}_n$ can be computed by an adaptation of a density level-set estimation algorithm \citep{H87} which we state below. Note that if a closed convex set $S^\star$ minimizes $\M_n$, the convex hull of $\{ X_i :  X_i \in S^\star, 1\leq i \leq n \}$ also minimizes $\M_n$. Hence, it suffices to reduce our search to convex polygons whose vertices belong to the set of $X_i$'s. 
There could be $2^n$ such polygons. So, an exhaustive search is computationally expensive.

{\it Computing the estimate.} We first find the optimal polygon (the convex polygon which minimizes $\M_n$) for each choice of $X$ as its leftmost vertex.  We use the following notation.  Let this particular $X$ be numbered 1, and let the $X_i$'s not to its left be numbered $2, 3, . . . , r$. The axes are shifted so that 1 is at the origin and the coordinates of point $i$ are denoted by $z_i$. The line segment $a z_i + (1 - a) z_j$, $(0 \leq a \leq  1)$ is written as $[i, j]$. Assume that $1, . . ., r$ are ordered so that the segments $[1, i]$ move counterclockwise as $i$ increases and so that $i \leq  j$ if $i \in [1, j]$. Polygons will be built up from triangles for $1 < i < j \leq  r$; $\Delta_{i j}$ is the convex hull of $(1, i, j)$ excluding $[1, i]$. Note that the segment $[1, i]$ is excluded from $\Delta_{i j}$ in order to combine triangles without overlap. The quadrilateral with vertices at $1, i, j, k$ for $i < j < k \leq r$ is convex if 
$$ D_{ijk} = \left| \begin{array}{cc}
z_i' & 1 \\
z_j' & 1 \\
z_k' & 1 \\	
\end{array} \right| \geq 0 $$

\begin{figure}
	\centering
		\includegraphics[width=0.50\textwidth]{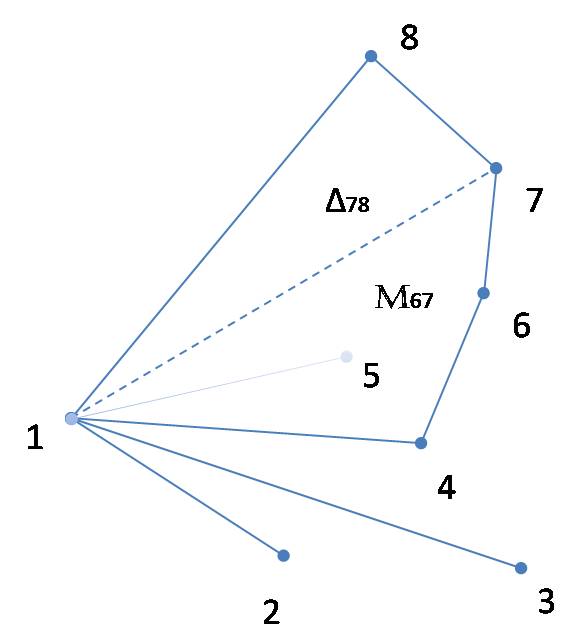}
	\caption{Notation for constructing the convex set estimate. An arbitrary vertex is numbered $1$, and those not to its left are numbered $2, 3, \ldots,8$ in a counterclockwise manner. The triangle $\Delta_{78}$ excludes the line segment $[1, 7]$. The optimal polygon (with measure $\M_{67}$) with successive vertices 6, 7 and 1 is depicted as the convex polygon with  vertices 1,4,6 and 7.}	\label{fig:h_algo0}  
\end{figure}
\begin{figure}
	\centering
		\includegraphics[width=0.50\textwidth]{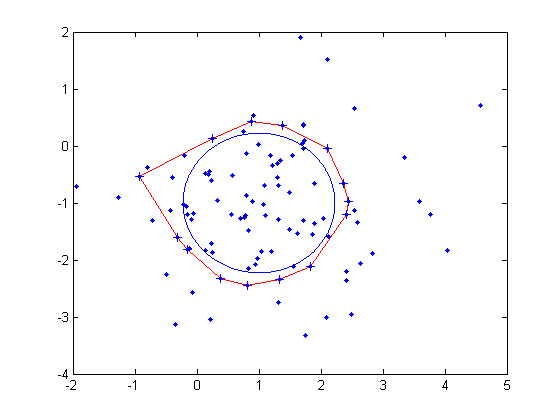}
	\caption{An illustration of the procedure in the dose-response setting with $m=10$ and $n=100$. The set $S_0$ is a circle centered at $(1,-1)$ with radius $1$.} 	\label{fig:h_algo1} 
\end{figure}

Let $\M_{1j}$ be the value of $\M_n$ on the line segment $[1, j]$. Further, for $1 < j < k \leq r$, let $\M_{jk}$ denote the minimum value of $\M_n$ among closed convex polygons with successive counterclockwise vertices $j, k$ and  $1$. Note that all such convex polygons contain the triangle 
$\Delta_{jk}$ and hence, $\M_n(\Delta_{jk})$, $\M_n$ measure of $\Delta_{jk}$, is a common contributing term to the $\M_n$ measure of all such polygons. This simple fact forms the basis of the algorithm.
It can be shown that 
\be
\M_{jk} = \M_{i^*j} + \M_n(\Delta_{jk}),
\label{c5eq:mrelation}
\ee where $i^* = I(k, j)$ is chosen to minimize $\M_{ij}$ over vertices $i$ with $i < j$, $D_{ijk } \geq  0$, i.e,
\be
 i^* = I(k, j) = \mathop{\mbox{argmin}}_{i: i<j, D_{ijk}>0} \M_{ij}.
\label{c5eq:irelation}
\ee 
Note that $i^*$ could possibly be $1$, in which case $\M_{jk}$ is simply the $\M_n$ measure of the triangle formed by $j$, $k$ and $1$ (including the contribution of line segment $[1,j]$).

One way to construct an optimal polygon with leftmost vertex $1$ is to find the minimum among $\M_{jk}$, $1\leq j < k$, where $\M_{jk}$'s are computed recursively using \ref{c5eq:mrelation} and \ref{c5eq:irelation}. 
Hence, one optimal polygon with leftmost vertex $1$ has vertices $i_l, i_2, \ldots, i_s = 1$, where either $s = 1$ or $\M_{i_2i_1} = \min_{1\leq j < k}  \M_{jk}$, $i_3 = I(i_1, i_2),$ $i_4 = I(i_2, i_3)$, \ldots, $1 = i_s= I(i_{s-2}, i_{s-1})$. 
Once this is done for each choice of $X$ as the leftmost vertex, the final estimate $\hat{S}_n$ is simply the one with the minimum $\M_n$ value among these $n$ constructed polygons. 

There are minor modifications to this algorithm which reduce the over-all implementation to 
$O(n^3)$ computations; see \citet[Section 3]{H87} for more details. 

\subsection{Regression Setting}\label{regset}
Consider a data generating model of the form 
\begin{eqnarray}
Y_{kl} = \mu(x_{kl}) + \epsilon_{kl}, \nonumber
\end{eqnarray}
with  $x_{kl} = ( u_k, v_l),$ $u_k = k/m$, $v_l = l/m$, $k, l \in \{ 1, \ldots, m \}$. The total number of observations is thus $n = m^2$. The errors $\epsilon_{kl}$s are independent with mean 0 and variance $\sigma^2_0$. Here, $\mu$ is as defined earlier and we seek to estimate $S_0 = \mu^{-1}(0)$. 

As earlier, we test the null hypothesis $H_{0,x}: \mu(x) = \tau_0$ against the alternative $H_{1,x}: \mu(x) > \tau_0$ at each level $x$ and use the resulting $p$-values to construct an estimate of the set $S_0$. For this, let $$\hat{\mu} (x) =  \frac{1}{n h_n^2 } {\sum_{k,l} {{Y}_{kl} K\left( \frac{{x- x_{kl}}}{h_n}\right)}}$$ 
 denote the estimator of $\mu$, with $K$ being a probability density (kernel) on $\R^2$ and $h_n$ the smoothing bandwidth. We take $h_n= h_0 n^{-\beta}$ for $\beta < 1/2$ and $K$ to be the $2$-fold product of a symmetric one-dimensional compact kernel, i.e., $K(x_1, x_2) =  K_0 (x_1) K_0(x_2)$, where $K_0$ is a symmetric probability density on $\R$ with $K_0(x) = 0$ for $|x| \geq L_0$.   

The statistic $T(x) = \sqrt{n h_n^2} (\hat{\mu}(x) - \tau_0) $ converges in distribution to a mean zero normal random variable with variance $\Sigma^2 = \sigma^2_0 \int_{u \in \R^2}{K^2(u) d(u)}$, when $x \in S_0$ and goes to $\infty$ when $x \notin S_0$. Hence, the non-normalized $p$-values  for testing $H_{0,x}$ against $H_{1,x}$ using $T$ can then be constructed as:
\begin{equation}
{p}_n(x) = 1 - \Phi\left(\sqrt{n h_n^2} (\hat{\mu}(x) - \hat{\tau} )\right), \nonumber
\end{equation} 
where $\hat{\tau}$ is a suitable estimate of $\tau_0$. 
These $p$-values asymptotically have mean 1/2 for $x \in S_0$ and converge to zero when $x \notin S_0$. Hence, as in Section \ref{doseset}, we  can estimate $S_0$ by minimizing
\begin{eqnarray}\label{c5eq:critrn}
\M_n(S) & = & \frac{1}{n}\sum_{k,l: x_{kl} \in \mathcal{I}_n}\left\{\Phi\left(\sqrt{n h_n^2} (\hat{\mu}(x_{kl} )- \hat{\tau}) \right) - \frac{3}{4}\right\} 1_{S}(x_{kl}) \\
& = & \frac{1}{n} \sum_{k,l: x_{kl} \in \mathcal{I}_n} \widetilde{W}_{kl} 1_{S}(x_{kl}) \nonumber
\end{eqnarray}
with $\widetilde{W}_{kl} = \Phi\left(\sqrt{n h_n^2} \hat{\mu}(x_{kl}) \right) - \gamma $ and $ \gamma = {3}/{4}$.  To avoid the bad behavior of the kernel estimator at the boundary, the sums are restricted to design points in $\mathcal{I}_n = [L_0 h_n, 1- L_0 h_n]^2$.  With $\mathcal{S}$ being the class of \emph{closed convex} subsets of $[0,1]^2$  as defined earlier, let 
 $\hat{S}_n = \mathop{\mbox{argmin}}_{S \in \mathcal S} \M_n (S)$. 

The estimate can be computed using the same algorithm as stated in Section 2.1.

\section{Notations and Assumptions}\label{sec:note}
We adhere to the setup of Sections \ref{doseset} and \ref{regset}, i.e., we assume the errors to be independent and  homoscedastic and consider random and fixed designs respectively for the dose-response and regression settings. A fixed design in the regression setting provides a simpler platform to illustrate the main techniques. In particular, it allows us to treat the kernel estimates as an $\tilde{m}$--dependent random field (where $\tilde{m}$ is specified later) which facilitates obtaining probability bounds on our estimate; see Section \ref{regrs2}. Also, a random design in the dose--response setting permits the use of empirical process techniques developed for i.i.d. data ($(\bar{Y}_i, X_i)$'s are i.i.d.). However, we note here that the dose-response model in a fixed (uniform) design setting can be addressed by taking an approach similar (and in fact, simpler due to the absence of smoothing) to that for the regression setting. 
The results on the rate of convergence of our estimate of $S_0$ are identical for the random design and the fixed uniform design dose-response
models.  

Let $\lambda$ denote the Lebesgue measure. The precision of the estimates is measured using the metrics
$$	 d_F(S_1, S_2)  = F( S_1 \Delta S_2) \mbox{ and }  \ d(S_1, S_2)  = \lambda( S_1 \Delta S_2)   $$
for the dose--response and the regression settings respectively. The two metrics arise naturally in their respective settings as $X_i$'s have distribution $F$ (in the dose--response setting) and the empirical distribution of  the grid points in the regression setting converges to the Uniform distribution on $[0,1]^2$.

For simplicity, we start assuming $\tau_0$ to be known. It can be shown that our results 
extend to cases where we impute a $\sqrt{mn}$ (dose-response)/ $\sqrt{n }$ (regression) estimate of $\tau$ (more on this in Section \ref{sec:extntau2}).
We summarize the assumptions below:
\begin{enumerate}
	\item The function $\mu$ is continuous on $[0,1]^2$. For the standard regression setting, we additionally assume that $\mu$ is Lipschitz continuous of order 1 .
\item The function $\mu$ is $p$-regular at $\partial S_0$, i.e., for some $\kappa_0, C_0 >0$ and for all $x \notin S_0$ such that $\rho(x, S_0) < \kappa_0$, 
\be
C_0 \rho(x, S_0)^p \leq \mu(x) - \tau_0 
\label{c5eq:smtmu}
\ee
Here $\rho$ is the $\ell_\infty$ metric in $\R^2$ (for convenience).
	\item $S_0 = \mu^{-1}(\tau_0)$ is convex. For some $\epsilon_0 >0$, ,  $S_0 \subset [\epsilon_0, 1-\epsilon_0]^2$ and $\lambda(S_0)>0$. 
	\item The design density $f$ for the dose-response setting  is assumed to be continuous and positive on $[0, 1]^2$.
\item Assumptions on the kernel $K(x)= K_0(x_1) K_0(x_2)$, $x = (x_1, x_2)$, for the standard regression setting:
\begin{enumerate}
	\item $K_0$ is a symmetric probability density.
	\item $K_0$ is compactly supported, i.e., $K_0(x) =0 $ when $|x| \geq L_0$, for some $L_0 >0$.
	\item $K$ is Lipschitz continuous of order 1.
\end{enumerate}

\end{enumerate}
Note that by the uniform continuity of $\mu$ and compactness of $[0,1]^2$, $\inf\{ \mu(x): \rho(x, S_0) \geq \kappa_0 \} > \tau_0$.
For a fixed $p$, $\tau_0$, $\kappa_0$, $\delta_0 >0$, we denote the class of functions $\mu$ satisfying assumptions 1, 2, 3 and 
\be
\inf\{ \mu(x): \rho(x, S_0) \geq \kappa_0 \} - \tau_0 > \delta_0
\label{c5eq:sepcondn}
\ee
 by $\mathcal{F}_p = \mathcal{F}_p(p,\tau_0, \kappa_0, \delta_0)$.
\begin{rmk}
It can be readily seen that if the regularity assumption in \eqref{c5eq:smtmu} holds for a particular $p$, it also holds for any $\tilde{p} > p$ as well. We assume that we are working with the smallest $p$ such that \eqref{c5eq:smtmu} is satisfied (the set of values $\tilde{p}$ such that \eqref{c5eq:smtmu} holds for a fixed $\mu$, $C_0$ and $\kappa_0$ is a closed set and is bounded from below whenever it is non-empty). In level-sets estimation theory, analogous two-sided conditions of the form
$$C_0 \rho(x, S_0)^p \leq |\mu(x) - \tau_0| < C_1 \rho(x, S_0)^p$$
are typically assumed (see \citet[Assumptions (4) and (4')]{T97},  \citet[Assumption (4)]{C97}). This stronger condition restricts the choice of $p$. However, we note here that the left inequality plays a more significant role  as it provides a lower bound on the amount by which $\mu(x)$ differs from $\tau_0$ in the vicinity of $\partial S_0$. Some results in a density level-set estimation problem with a slightly weaker analogue of the left inequality can be found in \cite{P95}. The upper bound (right inequality) is seen to be useful for establishing adaptive properties of certain density level-set estimates \citep{SNC09}.
\end{rmk}

\section{Consistency and Rate of Convergence}
\subsection{Dose-response setting}\label{dsrsp2}
%
As $\tau_0$ is known, we take $\tau_0 =0$ without loss of generality. Recall that $\M_n (S) = \P_n \left\{\Phi\left(\sqrt{m} \bar{Y} \right)- \gamma\right\} 1_S(X) .$ 
Let $P_{m}$ denote the measure induced by $(\bar{Y}, X)$ and
\be
M_{m}(S) = P_{m} \left[\left\{\Phi\left(\sqrt{m}\bar{Y}\right) - \gamma\right\}1_S(X)\right]. \nonumber
\ee
The process $M_{m}$ acts as a population criterion function and can be simplified as follows.
Let
\be \label{c5eq:Z_1n}
Z_{1 m}  = \frac{1}{\sqrt{m} \sigma_0}\sum_{j=1}^m \epsilon_{1j}
\ee
and $Z_0$ be a standard normal random variable independent of $Z_{1 m}$s. Then
\begin{eqnarray}
E\left[\left. \Phi \left( \sqrt{m} {\bar{Y}_1 } \right) \right| X_1 = x   \right]  & = & E\left[\Phi \left( \sqrt{m} {\mu}(x)  + \sigma_0 Z_{1 m} \right) \right]  \nonumber\\
& =& E \left[ E \left[ \left. 1 \left( Z_0 < \sqrt{m} {\mu}(x)+ \sigma_0 Z_{1 m} \right) \right| Z_{1 m} \right] \right] \nonumber \\
& =& P \left[ \frac{Z_0 - \sigma_0 Z_{1 m} }{\sqrt{1 + \sigma_0^2}} < \frac{\sqrt{m} {\mu}(x)}{{\sqrt{1 + \sigma_0^2}}} \right] = \Phi_m \left( \frac{\sqrt{m} {\mu}(x)}{{\sqrt{1 + \sigma_0^2}}} \right), \nonumber
\end{eqnarray}
where $\Phi_m$ denotes the distribution function of $(Z_0 - \sigma_0 Z_{1 m})/\sqrt{1 + \sigma_0^2}$. By P{\'o}lya's theorem, $\Phi_m$ converges uniformly to $\Phi$ as $m \rightarrow \infty$. Hence, it can be seen that
$$ \lim_{m \rightarrow \infty } E\left[\left. \Phi \left( \sqrt{m} {\bar{Y}_1 } \right) \right| X_1 = x   \right] = \frac{1}{2} 1_{S_0}(x) +  1_{S_0^c}(x).$$
By the Dominated Convergence Theorem, $M_m(S)$ converges to $M(S)$, where 
\begin{eqnarray}
 M(S) = M_F(S) & = & \int_S{\left(\frac{1}{2} 1_{S_0}(x) +  1_{S_0^c}(x) - \gamma\right)} F(dx) \nonumber\\
& = & (1/2 - \gamma) F(S_0 \cap S) + (1- \gamma) F(S_0^c \cap S).
\label{c5eq:constfn}
\end{eqnarray}
Note that $S_0$ minimizes the limiting criterion function $M(S)$. 
An application of the argmin continuous mapping theorem \citep[Theorem 3.2.2]{VW96} yields the following result on the consistency of $\hat{S}_n$
\begin{thm} \label{c5th:constmn}
Assume $S_0$ to be a closed convex set and the unique minimizer of $M(S)$. 
Then $\sup_{S \in \mathcal{S}}|\M_n(S) - M(S) | $ and  $d_F(\hat{S}_n, S_0)$ converge in outer probability to zero for any $\gamma \in (0.5,1)$.
\end{thm}
\begin{rmk}\label{rmk:const} We end up proving a stronger result. The consistency is established in terms of the Hausdorff metric which implies consistency with respect to $d_F$. Moreover, we do not require $m$ to grow as $m_0 n^\beta$, $\beta >0$ for consistency. The  condition $\min(m, n) \rightarrow \infty$ suffices. Also, the result extends to higher dimensions as well, i.e., when $\mu$ is a function from $[0,1]^d \mapsto \R$ and $S_0= \mu^{-1}(0)$ is a closed convex subset of $[0,1]^d$, then the analogous estimate is consistent.  However, an efficient way to compute the estimate is not immediate. 
\end{rmk}
The proof is given in Section \ref{c5pf:constmn} of the Appendix.
\newline
\newline
We now proceed to deducing the rate of convergence of $d_F(\hat{S}_n,S_0)$.  For this, we study how small the difference $(\M_n - {M})$ is and how $M$ behaves in the vicinity of ${S}_0$. 
We split the difference $(\M_n - {M})$ into $(\M_n - {M}_m)$ and $({M}_m - {M})$ and study them separately. The term $\M_n - M_m$ involves an empirical average of centered random variables, efficient bounds on which are derived using empirical process inequalities. 
We start with establishing a bound on the non-random term $({M}_m - {M})$ in the vicinity of $S_0$. To this end, we first state a fact that gets frequently used in the proofs that follow. 
\newline
{\bf Fact:} For any $\delta >0$, let $S^\delta = \{ x: \rho(x, S) < \delta \}$ and $_\delta S = \{ x : \rho(x, S_0^c) \geq \delta\}$ denote the $\delta$-fattening and $\delta$-thinning of the set $S$. There exists a constant $c_0>0$ such that for any $S \in \mathcal{S}$,
\be
\lambda (S^\delta \backslash _\delta S) \leq c_0 \delta \mbox{ and consequently, } F(S^\delta \backslash _\delta S) \leq \tilde{c}_0 \delta,
\label{eq:thinbnd}
\ee
with $\tilde{c}_0 = \|f\|_\infty c_0$ ($\|f\|_\infty < \infty$, by Assumption 4). For a proof of the above, see, for example, \citet[pp. 62--63]{ddly84}. 

\begin{lem} \label{lm:constboundmn}
For any $\delta >0$, $a_n \downarrow 0$ and $S \in \mathcal{S}$ such that $F(S \triangle S_0) < \delta$, 
\begin{eqnarray*}
{\left| (M_m - {M})(S) -  (M_m - {M})(S_0) \right|} &\leq& |\Phi_m(0) - 1/2| \delta + \min( \tilde{c}_0 a_n, \delta) \\
& & +  \left|\Phi_m\left(\frac{C_0 \sqrt{m} a_n ^p}{ \sqrt{1 + \sigma_0^2}} \right) - 1\right| \delta \\
& &+  \left|\Phi_m\left(\frac{\sqrt{m} \delta_0}{ \sqrt{1 + \sigma_0^2}} \right) - 1\right|\delta. 
\end{eqnarray*}
\end{lem}
{\it Proof.} Note  that
\begin{eqnarray*}
M_m(S) - M_m (S_0) & = & P_m \left[\left\{\Phi_m\left(\frac{\sqrt{m} {\mu}(x)}{\sqrt{1 + \sigma_0^2}} \right) - \gamma\right\} \left\{1_{S}(x) - 1_{S_0}(x) \right\}\right] \mbox{ and } \\
{M}(S) - {M} (S_0) & = & \int \left\{(1/2) 1_{S_0}(x) + 1_{S_0^c}(x)   - \gamma\right\} \left\{1_{S}(x) - 1_{S_0}(x) \right\} F(dx).
\end{eqnarray*}
Hence, the expression $\left| (M_m - {M})(S) -  (M_m - {M})(S_0) \right|$ is bounded by
 \begin{eqnarray}
& & \int_{x \in (S_0 \cap S) } \left|\Phi_m \left(0 \right) - \frac{1}{2}\right| F(dx)  +  \int_{x \in (S_0^c \cap S) } \left|\Phi_m \left(\frac{\sqrt{m} {\mu}(x)}{\sqrt{1 + \sigma_0^2}} \right) - 1\right| F(dx) . 
\label{c5eq:frstbndmn}
\end{eqnarray}
Note that the first term is bounded by $|\Phi_m(0) - 1/2| \delta$.  Further, let $S_n = \{ x: \rho(x, S_0) \geq a_n \}$. 
Using \eqref{eq:thinbnd}, $F(S_n^c \backslash S_0) \leq \tilde{c}_0 a_n$. Also, as $a_n \downarrow 0$, $a_n < \kappa_0$ for sufficiently large $n$. Thus, for $x \in S_n$, 
$$\mu(x) \geq \min(\rho(x, S_0)^p, \delta_0) \geq \min(a_n^p, \delta_0), $$ using \eqref{c5eq:smtmu} and \eqref{c5eq:sepcondn}. Hence, the second sum in \eqref{c5eq:frstbndmn} is bounded by
{ \begin{eqnarray*}
\lefteqn{F(S_n^c \backslash S_0) + \int_{x \in (S_n \cap S) }\left|\Phi_m \left(\frac{\sqrt{m} {\mu}(x)}{\sqrt{1 + \sigma_0^2}} \right) - 1\right| F(dx) \leq \min(\tilde{c}_0 a_n, \delta) } \\
& & + \int_{x \in (S_n \cap S) } \left\{\left|\Phi_m\left(\frac{C_0 \sqrt{m} a_n^{p}}{\sqrt{1 + \sigma_0^2}} \right) - 1\right| + \left|\Phi_m\left(\frac{\sqrt{m} \delta_0}{\sqrt{1 + \sigma_0^2}} \right) - 1\right| \right\}F(dx) . 
\end{eqnarray*}}
As $F(S_n \cap S) < \delta$, we get the result.
\qed
\newline
\newline
To control $\M_n - M_m$, we rely on a version of Theorem 5.11 of \cite{V00}. The result in its original form is slightly general. In their notation, it involves a bound on a special metric $\rho_K(\cdot)$  (see \citet[equation 5.23]{V00}) which, in light of Lemma 5.8 of \cite{V00}, can be controlled by bounding the $L_2$-norm in the case of bounded random variables. This yields the consequence stated below. Here, $H_B$ denotes the entropy with respect to bracketing numbers.
\begin{thm}\label{thm:vdgrbrk}
Let $\mathcal{G}$ be a class of functions such that $\sup_{g \in \mathcal{G}} \|g\|_{\infty} \leq 1 $. For some universal constant $C>0$, let $C_2, C_3, R$ and $N >0$ satisfy the following conditions:
\begin{eqnarray*}
R & \geq & \sup_{g \in \mathcal{G}} \|g\|_{L_2(P)}, \\ 
N &\geq & C_2 \int_0^R H^{1/2}_B(u, \mathcal{G}, L_2(P)) du \vee R \\
C_2^2 & \geq & C^2 (C_3 + 1) \mbox{ and } \\
N & \leq & C_3 \sqrt{n} R^2.
\end{eqnarray*}
Then
$$ P^*\left[\sup_{g \in \mathcal{G}}|\G_n(g)| > N \right] \leq C \exp\left[\frac{- N^2}{C^2 (C_3 + 1)R^2}\right],$$
where  $P^*$ denotes the outer probability.
\end{thm}

We have the following theorem on the rate of convergence of $\hat{S}_n$. 
\begin{thm}\label{c5prp:ratemu}
For any $\beta >0$, 
$$P^*\left(d_F(\hat{S}_n, {S}_0) > \delta_n \right) \rightarrow 0$$
for $\delta_n = K_1 \max \{n^{-2/3}, m^{-1/(2p)}\}$, where $K_1>0$ is some constant. 
\end{thm}
{\it Proof.} 
Let $k_n$ be the smallest integer such that $2^{k_n+1} \delta_n \geq 1$.
For $0 \leq  k \leq k_n$, let $\mathcal{S}_{n,k} = \left\{S : S \in \mathcal{S},\ 2^k \delta_n < d_F(S, {S}_0 ) \leq  2^{k+1} \delta_n\right\}$.
As $\hat{S}_n$ is the minimizer for $\M_n$, 
\begin{eqnarray*}
{ P^* \left(d_F(\hat{S}_n, {S}_0) > \delta_n\right) }  &\leq& \sum_{k=0}^{k_n} P^*\left( \inf_{A \in \mathcal{S}_{n,k}}\M_n(A) - \M_n({S}_0) \leq 0 \right) . 
\end{eqnarray*}
The sum on the right side is bounded by 
\begin{equation} 
\sum_{k=0}^{k_n} P^* \left(\sup_{A \in \mathcal{S}_{n,k}} \left| (\M_n - M)(S) -  (\M_n - M)(S_0) \right| > \inf_{A \in \mathcal{S}_{n,k}} \left({M}(S) - {M}({S}_0) \right)  \right). 
\label{c5eq:shellmn}
\end{equation}
For $c(\gamma) = \min (\gamma - 1/2, 1 - \gamma)>0$, 
$$ M(S)  - M(S_0) = ( \gamma - 1/2) (F(S_0) - F(S_0 \cap S)) + (1- \gamma) F(S_0^c \cap S) \geq c(\gamma) F(S \triangle S_0),$$
and hence, \eqref{c5eq:shellmn} is bounded by
\begin{eqnarray}
\lefteqn{\sum_{k=0}^{k_n} P^* \left(\sup_{A \in \mathcal{S}_{n,k}} \left| (\M_n - {M}_m)(S) -  (\M_n - {M}_m)(S_0) \right| > c(\gamma) 2^{k-1} \delta_n \right) } \nonumber \\
&& + \sum_{k=0}^{k_n} P^* \left( \sup_{A \in \mathcal{S}_{n,k}} \left| ({M}_m - {M})(S) -  ({M}_m - {M})(S_0) \right| \geq c(\gamma) 2^{k-1} \delta_n   \right). 
\label{c5eq:simpmn}
\end{eqnarray}

Note that $M_m - M$ is a non-random process and hence, each term in the second sum is either 0 or 1. 
We now show that the second sum in the above display is eventually zero. For this, we apply Lemma \ref{lm:constboundmn}. Note that
\begin{eqnarray}
\lefteqn{\sup_{A \in \mathcal{S}_{n,k}} \left| ({M}_m - {M})(S) -  ({M}_m - {M})(S_0) \right| }\nonumber\\
&\leq  &  |\Phi_m(0) - 1/2| 2^{k+1}\delta_n +\min(\tilde{c}_0 a_n, 2^{k+1}\delta_n) \nonumber\\
&& + \left|\Phi_m\left(\frac{C_0 \sqrt{m} a_n ^p}{ \sqrt{1 + \sigma_0^2}} \right) - 1\right| 2^{k+1}\delta_n + \left|\Phi_m\left(\frac{\sqrt{m} \delta_0}{ \sqrt{1 + \sigma_0^2}} \right) - 1\right| 2^{k+1}\delta_n \nonumber\\
& \leq  &  4\left[|\Phi_m(0) - 1/2| + \left|\Phi_m\left(\frac{\sqrt{m} \delta_0}{ \sqrt{1 + \sigma_0^2}} \right) - 1\right|\right] 2^{k-1} \delta_n \nonumber\\
& & + \left[\frac{2\tilde{c}_0 a_n}{\delta_n}+ 4 \left|\Phi_m\left(\frac{C_0 \sqrt{m} a_n ^p}{ \sqrt{1 + \sigma_0^2}} \right) - 1\right| \right] 2^{k-1} \delta_n. 
\label{c5eq:constbound}
\end{eqnarray}
Hence, it suffices to show that the coefficient of $2^{k-1} \delta_n$ in the above expression is smaller than $c(\gamma)$. To this end, fix $0 < \eta < c(\gamma)/8$. For large $m$, $$|\Phi_m(0) - 1/2| + \left|\Phi_m\left({\sqrt{m} \delta_0}/{ \sqrt{1 + \sigma_0^2}} \right) - 1\right| \leq \eta. $$ Choose $c_\eta$ such that $a_n= c_\eta m^{-1/(2p)} > \left[\Phi_m^{-1}(1 - \eta)\sqrt{1 + \sigma_0^2}/ ( C_0 \sqrt{m}) \right]^{1/p}$. For large $n$, the coefficient of $2^{k-1}\delta_n$ in \eqref{c5eq:constbound} is then bounded by
$$ 8\eta + \frac{\tilde{c}_0 c_\eta}{K_1} < c(\gamma),$$ 
for $K_1 > (\tilde{c}_0 c_\eta)/ (c(\gamma) - 8 \eta) $. Hence, each term in the second sum of \eqref{c5eq:simpmn} is zero for a suitably large choice of the constant $K_1$. 
Note that the first term in \eqref{c5eq:simpmn} can be written as 
\begin{eqnarray}
\sum_{k=0}^{k_n} P^*\left(\sup_{A \in \mathcal{S}_{n,k}} \left| \G_n g_m(\bar{Y}) 1_{A \triangle S_0}(X) \right| > c(\gamma){2^{k-1} \delta_n \sqrt{n} }\right),
\label{c5eq:sumineqmn}
\end{eqnarray}
where  
 $g_m (y) = \Phi\left(\sqrt{m} y\right) - \gamma$. We are now in a position to apply Theorem \ref{thm:vdgrbrk} to each term of \eqref{c5eq:sumineqmn}. In the setup of Theorem \ref{thm:vdgrbrk}, $N = c(\gamma) 2^{k-1} \delta_n \sqrt{n} $. The concerned class of functions is $\mathcal{G}_{n,k} = \{ g_m(\bar{Y}) 1_B(X) : B = A \triangle S_0, B \in \mathcal{S}_{n,k} \}$. Note that $\| g_m 1_B \|_{L_2(P)} \leq [E 1_B(X)]^{1/2} \leq  (2^{k+1} \delta_n)^{1/2}$. So we can pick $R = R_{n,k} =  (2^{k+1} \delta_n)^{1/2}$. As $\mathcal{S}_{n,k} \subset \mathcal{S}$, $N_{[ \ ]} (u, \{A\triangle S_0: A \in \mathcal{S}_{n,k} \}, L_2(P))  \leq (N_{[ \ ]} (u, \mathcal{S}, L_2(P)))^2$ for any $u>0$. Also, starting with a bracket $[f_L, f_U]$ for $\{A\triangle S_0: A \in \mathcal{S}_{n,k} \}$ containing $B$ with $\|f_U -f_L\|_{L_2(P)} \leq u$,  we can obtain brackets for the class $\mathcal{G}_{n,k}$ using the inequality
$$\Phi\left(\sqrt{m} y\right)f_L - \gamma f_U \leq g_m(y) 1_B(x) \leq \Phi\left(\sqrt{m} y\right)f_U - \gamma f_L. $$
As $\|g_m\|_{\infty} \leq 1$, 
$$\|(\Phi\left(\sqrt{m} y\right)f_U - \gamma f_L) -  (\Phi\left(\sqrt{m} y\right)f_L - \gamma f_U) \|_{L_2(P)} \leq u.$$
Hence, $H_B (u, \mathcal{G}_{n,k}, L_2(P)) \leq H_B (u, \mathcal{S}, L_2(P))$. 
  Using the fact that in dimension $d$, $H_B(u, \mathcal{S}, L_2(P)) =  \log(N_{ [ \ ] } (u, \mathcal{S}, L_2(P))) \leq A_0 u^{-(d-1)}$ for $d \geq 2$ (see \cite{B76}), we get 
$$ H_B (u, \mathcal{G}_{n,k}, L_2(P)) \leq A_0 u^{-1} $$
for some constant $A_0 >0$ (depending only on the design distribution). The conditions of Theorem \ref{thm:vdgrbrk} then translate to 
\begin{eqnarray*}
2^{k-1} c(\gamma) \delta_n \sqrt{n}  &\geq & 2 C_2 \max(A_0,1) (2^{k+1} \delta_n)^{1/4} \\
C_2^2 & \geq & C^2 (C_3 + 1) \mbox{ and } \\
c(\gamma)2^{k-1} \delta_n \sqrt{n}   & \leq & C_3 \sqrt{n} 2^{k+1} \delta_n.
\end{eqnarray*}
It can be seen that for  $K_1 \geq 2^9 (C_2 \max(A_0,1)/c(\gamma))^{4/3 } $, $C_3 =c(\gamma)/4$ and $C_2 = \sqrt{5} C/2$,  these conditions are satisfied, and hence, we can bound \eqref{c5eq:sumineqmn} by
\begin{eqnarray*}
 \sum_{k=0}^{k_n} C \exp\left\{\frac{- 2^{k-3} c^2(\gamma) \delta_n  n }{C^2(C_3+1)}\right\}\\
 \end{eqnarray*}
As $\delta_n \gtrsim  n^{-2/3}$ (the symbol $\gtrsim$ is used to denote the corresponding $\geq$ inequality holding up to some finite positive constant), the term $\delta_n  n $ diverges to $\infty$ as $n \rightarrow \infty$. Hence, the above display converges to zero. 
This completes the proof.
\qed
\begin{rmk}
The result obviously holds for values of $\delta_n$ larger than the one prescribed above. Hence, it also gives consistency, though it requires $m$ to grow as $m_0 n^\beta$. In terms of the total budget, choosing $\beta = 4p/3$ corresponds to the optimal rate in which case $\delta_n$ is of the order $n^{-2/3}$ or $N^{-2/(4p+3)}$. To see this, we just set $n^{-2/3} = m^{-1/2p}$ and follow up the implications of this for $\beta$. The $N^{-2/(4p+3)}$ rate coincides with the minimax rate obtained for a related density level set problem in \citet[Theorem 2]{T97} (see also \citet[Theorem 3.7]{P95}).
\end{rmk}
Note that the bounds deduced for the two sums in \eqref{c5eq:simpmn} depend on $\mu$ only through $p$ and $\delta_0$, e.g., the exponential bounds from Theorem \ref{thm:vdgrbrk} depend on the class of functions only through their entropy and norm of the envelope which do not change with $\mu$.  Hence, we have the following result which is similar in flavor to the upper bounds deduced for level-set estimates in \cite{T97}.  
\begin{cor}\label{cor:expbnd} 
For the choice of $\delta_n$ given in Theorem \ref{c5prp:ratemu}, 
\be \label{c5eq:expbnd}
\lim \sup_{n \rightarrow \infty } \sup_{\mu \in \mathcal{F}_p} E^*_\mu \left[\delta_n^{-1} d(\hat{S}_n,S_0)\right] < \infty. 
\ee
Here, $E_\mu$ is the expectation with respect to the model with a particular $\mu \in \mathcal{F}_p$. The other features of the model such as error distribution and the design distribution do not change.
\end{cor}
{\it Proof.} Note that 
\begin{eqnarray*}
E^*_\mu \left[\delta_n^{-1} d(\hat{S}_n,S_0)\right] & \leq & 1 + \sum_{k\geq 0, 2^k \delta_n \leq 1} 2^{k+1}P^*\left( 2^k < \delta_n^{-1} d(\hat{S}_n, S_0) \leq 2^{k+1}\right) \\
&\leq& 1 + \sum_{k\geq 0, 2^k \delta_n \leq 1} 2^k P^*\left( \inf_{A \in \mathcal{S}_{n,k}}\M_n(A) - \M_n({S}_0) \leq 0 \right). 
\end{eqnarray*}
 The probabilities $P^*\left( \inf_{A \in \mathcal{S}_{n,k}}\M_n(A) - \M_n({S}_0) \leq 0 \right)$  can be bounded in an identical manner to that in the proof of the Theorem \ref{c5prp:ratemu} and hence, we get
\begin{eqnarray*}
\sup_{\mu \in \mathcal{F}_p} E^*_\mu \left[\delta_n^{-1} d(\hat{S}_n,S_0)\right] & \leq & 1 +  \sum_{k=0}^{k_n} C 2^{k+1} \exp\left\{\frac{- 2^{k-3} c^2(\gamma) \delta_n  n }{C^2(C_3+1)}\right\}.
\end{eqnarray*}
As $\delta_n n \rightarrow \infty$, the right side of the above is bounded and hence, we get the result.
\qed
 
%

\subsection{Regression Setting}\label{regrs2}
With $\tau_0 = 0$, recall that 
\begin{eqnarray}
\M_n(S) = \frac{1}{n}\sum_{k,l: x_{kl} \in \mathcal{I}_n}\left\{\Phi\left(\sqrt{n h_n^2} \hat{\mu}(x_{kl}) \right) - \gamma\right\} 1_{S}(x_{kl}). \nonumber
\end{eqnarray}
For any fixed $\gamma \in (1/2,1)$, it can be shown that $\hat{S}_n$ is consistent for $S_0$, i.e., $d(\hat{S}_n, S_0)$ converges in probability to zero.
\begin{thm}\label{c5th:constreg}
Assume $S_0$ to be a closed convex set and the unique minimizer of $M(S)$, where $$M(S)=  (1/2 - \gamma) \lambda (S_0 \cap S) + (1- \gamma) \lambda (S_0^c \cap S).$$
Then, $\sup_{S \in \mathcal{S}} |\M_n(S) - M(S)|$ converges in probability to zero and $\hat{S}_n$ is consistent for $S_0$ in the sense that $ d(\hat{S}, S_0)$ converges in probability to zero for any $\gamma \in (0.5,1)$.
\end{thm}
As was the case in the dose-response setting (see Remark \ref{rmk:const}), a more general result holds and is proved in Section \ref{c5pf:constreg} of the Appendix.

We now deduce a bound on the rate of convergence of $\hat{S}_n$ (for a fixed $\gamma \in (1/2,1)$).
%
We first consider the population equivalent of $\M_n$, given here by $\bar{M}_n (S) = E \{ \M_n (S) \} $ which can be simplified as follows. Let
\be
Z_{k l}  = \frac{1}{\sqrt{n h_n^2} } \sum_{k',l' : x_{kl}\in \mathcal{I}_n}^n \epsilon_{k'l'} K\left( \frac{x_{k l} - x_{k'l'}}{h_n}\right), \nonumber
\ee
 for $k = 1, \ldots, n$, and $Z_0$ be a standard normal random variable independent of $Z_{k l}$'s. For notational simplicity, $\sum_{k,l}$ (equivalently, $\sum_{k',l'}$) is used to denote a sum over the set $\{k,l : x_{kl} \in \mathcal{I}_n\}$ unless stated otherwise. Also, let
\be\label{c5eq:mubar}
\bar{\mu} (x) =  \frac{1}{\sqrt{n h_n^2} } {\sum_{k',l'} {{\mu}(x_{k'l'}) K\left( \frac{x- x_{k'l'}}{h_n}\right)}} \mbox{ and }\Sigma^2_n(x) = \frac{1}{n h_n^2 } {\sum_{k',l'} \sigma_0^2 K^2\left( \frac{x- x_{k'l'}}{h_n}\right)}.
\ee
Note that $\sqrt{n h_n^2} \hat{\mu} (x_{kl}) = \sqrt{n h_n^2} \bar{\mu} (x_{kl}) + Z_{kl}$ and $\mbox{Var} (Z_{kl}) = \Sigma_n^2(x_{kl})$. 
We have
\begin{eqnarray}
E \left[ \Phi\left(\sqrt{n h_n^2} \hat{\mu} (x_{kl})\right) \right]   & = & E\left[\Phi\left(\sqrt{n h_n^2} \bar{\mu} (x_{kl}) + Z_{kl} \right)\right]\nonumber\\
& =& E \left[ 1 \left( Z_0  \leq \sqrt{n h_n^2} \bar{\mu} (x_{kl}) + Z_{kl} \right) \right] \nonumber \\
& =& \Phi_{kl,n}\left(\frac{\sqrt{nh_n^2}\bar{\mu}(x_{kl})}{\sqrt{1 + \Sigma^2_n(x_{kl})}}\right), \nonumber 
\end{eqnarray}
where $\Phi_{kl,n}$ denotes the distribution function of $\left({Z_0 - Z_{kl} }\right)/{\sqrt{1+\Sigma^2_n(x_{kl})}}$. For $x_{kl} \in \mathcal{I}_n$,  $\Sigma^2_n(x_{kl})$ and $\Phi_{kl,n}$ do not vary with $k$ and $l$ and are  denoted by $\tilde{\Sigma}^2_n$ and $\tilde{\Phi}_n$, respectively, for notational convenience. 
 We get  
\begin{eqnarray}
& & \bar{M}_n (S) = E \{ \M_n (S) \} = \frac{1}{n}\sum_{k,l}\left\{\tilde{\Phi}_n\left(\frac{\sqrt{n h_n^2} \bar{\mu}(x_{kl})}{\sqrt{1 + \tilde{\Sigma}^2_n}} \right) - \gamma\right\} 1_{S}(x_{kl}).
\label{c5eq:means}
\end{eqnarray} 
Also, for $x_{kl} \in \mathcal{I}_n$, any $\eta > 0$ and sufficiently large $n$, 
{\small $$
\frac{1}{{n h_n^2} \tilde{\Sigma}^2_n} \sum_{\substack{k',l' \\ \rho(x_{kl}, x_{k'l'}) \leq L_0 m h_n} }  E\ \left[ \epsilon^2_{k'l'} K^2\left( \frac{x_{kl}- x_{k'l'}}{h_n}\right)
1 \left(\frac{|\epsilon_{k'l'}| K\left( ({x_{kl}- x_{k'l'}})/{h_n}\right) }{{\sqrt{n h_n^2}} \tilde{\Sigma}_n}  > \eta \right) \right] $$}	
is bounded by 
$$\frac{2 \left\lceil 2 L_0 m h_n  \right\rceil^2 \|K \|_\infty^2}{ n h_n^2 (\sigma^2_0 {\bar{K^2}})} E \left[ \epsilon_{11}^2 1 \left(\frac{2\|K \|_\infty}{{n h_n^2} (\sigma_0 \sqrt{\bar{K^2}})} |\epsilon_{11}|> \eta \right) \right],$$
which converges to zero. Hence, by Lindeberg--Feller central limit theorem, $Z_{kl}/\tilde{\Sigma}_n$ and consequently, $\tilde{\Phi}_n$ converge weakly to $\Phi$. Further, by P{\'o}lya's theorem, $\tilde{\Phi}_n$  converges uniformly to $\Phi$ as $n \rightarrow \infty$, a fact 
we use in the proof of Lemma \ref{lm:constbound}.

We now consider the distance $d(\hat{S}_n, S_0)$, the rate of convergence of which is driven by the behavior of  how small the difference $\M_n - {M}$ is and how $M$ behaves in the vicinity of ${S}_0$. 
As before, we split the difference $\M_n - {M}$ into $\M_n - \bar{M}_n$ and $\bar{M}_n - {M}$ and study them separately.
We first derive a bound on the distance between $\bar{M}_n$ and $M$.
\begin{lem} \label{lm:constbound}
There exist a positive constant $c_1$ such that for any $a_n \downarrow 0$ satisfying $a_n > 2 L_0 h_n$, $\delta>0$ and $\lambda(S \triangle S_0) < \delta$,  
\be
\begin{split}
\left| (\bar{M}_n - {M})(S) -  (\bar{M}_n - {M})(S_0) \right| &\leq |\tilde{\Phi}_n(0) - 1/2| \delta  + \min(c_0 a_n,\delta)  \\
&  + \left|\tilde{\Phi}_n\left(\frac{\sqrt{n h_n^2} C_0 (a_n - 2L_0 h_n )^p}{2 \sqrt{1 + \Sigma^2}} \right) - 1\right| \delta \\ 
& + \left|\tilde{\Phi}_n\left(\frac{\sqrt{n h_n^2} \delta_0}{2 \sqrt{1 + \Sigma^2}} \right) - 1\right| \delta + c_1 h_n. 
\label{c5eq:meandiff}
\end{split}
\ee 
%
\end{lem}
The proof of this lemma is available in Section \ref{cbapp} of the appendix. 
\newline
\newline
We next consider the term $ \M_n(S) - \bar{M}_n(S)$. With $\widetilde{W}_{kl}$s as defined in \eqref{c5eq:critrn}, let ${W}_{kl}  = \widetilde{W}_{kl}  - E \{ \widetilde{W}_{kl} \}$. Then $$ \M_n(S) - \bar{M}_n(S) =  \frac{1}{n} \sum_{k,l} {W}_{kl} 1_{S}(x_{kl}). $$
For notational ease, we define $W_{kl} \equiv 0$ whenever $x_{kl} \notin \mathcal{I}_n$. As the kernel $K$ is compactly supported, $W_{kl}$ is independent of all $W_{k'l'}$s except for those in the set  $\{ W_{k'l'} : (k',l') \in \{1,\ldots,m\}^2, \rho((k,l), (k',l') ) \leq 2 L_0 m h_n \}$. The cardinality of this set is at most $m' = 16L_0^2 n h_n^2$.  Hence, $\{ W_{kl} \}_{1 \leq k,l \leq m}$ is an $(\sqrt{m'}/2)$-dependent random field.  
For 
\be \label{c5eq:indices}
k_i = i + k \left\lceil \sqrt{m'}\right\rceil, \, l_j = j + l \left\lceil \sqrt{m'}\right\rceil \mbox{ and } r_{ij} = \sum_{\substack{k,l : 1 \leq k_i, l_j \leq m }} 1,
\ee
let 
\begin{eqnarray*}
\lefteqn{\tilde{d}_n(S_1, S_2) = \tilde{d}_n(1_{S_1}, 1_{S_2}) }\\
& =& \left\{\max_{1 \leq i, j \leq \left\lceil \sqrt{m'}\right\rceil} \left[\frac{1}{r_{ij}}\sum_{\substack{k,l : 1 \leq k_i, l_j \leq m }} (1_{S_1} (x_{k_i l_j}) - 1_{S_2} (x_{k_i l_j}) )^2 \right] \right\}^{1/2}
\end{eqnarray*}
and $ \| S \|_n = \| 1_S \|_n = \tilde{d}_n(S, \phi) $.
Then, the following relation holds.
\begin{lem}\label{lm:onedubineq}
For sufficiently large $n$, 
\be
{ P \left( \frac{1}{n}\left| \sum_{k,l} {W}_{kl} 1_{S}(x_{kl}) \right| \geq a \right) } \leq 2 \exp \left[-\frac{n a^2}{16 m'\|S\|_n^2 }\right]. \nonumber
\ee 
\end{lem}
The proof is given in Section \ref{c5pf:onedubineq} of the Appendix. In fact, such a result holds for general (bounded) $(\sqrt{m'}/2)$-dependent random fields $\{V_{kl}: 1 \leq k,l \leq m \}$ with $|V_{kl}| \leq 1$ and weights $g(x_{kl})$ (instead of $1_S (x_{kl})$'s) with $\tilde{d}_n(g_1, g_2) $ defined accordingly, as long as $n/m' \rightarrow \infty$, i.e., it can be shown that 
\be
{ P \left( \frac{1}{n}\left| \sum_{k,l} {V}_{kl} g(x_{kl}) \right| \geq a \right) } \leq 2 \exp \left[-\frac{n a^2}{16 m'\|g\|_n^2 }\right]. 
\label{c5eq:simpleineq1}
\ee 
Moreover, we can generalize the above to a probability bound on the supremum of an empirical process.
\begin{thm}\label{thm:vdgr}
Let $\mathcal{G}$ denote a class of weight functions $g: \{ x_{kl}: 1 \leq k,l \leq m \} \mapsto \R$ and $H$ denote the entropy of this class with respect to covering numbers and the metric $\tilde{d}_n$.  Assume  $\sup_{g \in \mathcal{G}} \| g\|_{n} \leq R$. Let $V_{kl}$s be random variables  with $|V_{kl}| \leq 1$ such that the inequality \eqref{c5eq:simpleineq1} holds for all $g \in \mathcal{G}$. Then, there exists a universal constant $C>0$ such that for all $ \delta_1 > \delta_2 \geq 0 $ satisfying
\be \label{c5eq:conditionch}
\sqrt{n/m'} (\delta_1 - \delta_2) \geq C \left(\int_{\delta_2/8}^R H^{1/2}(u, \mathcal{G}, \tilde{d}_n ) du \vee R \right), 
\ee
we have
$$ P^* \left[\sup_{g\in \mathcal{G}} \left| \frac{1}{n}\sum_{k,l} {V}_{kl} g(x_{kl})\right|\geq \delta_1 - \delta_2 \right] \leq C \exp \left[-\frac{n (\delta_1- \delta_2)^2}{C m' R^2}\right]. $$
\end{thm}
The above result states that the supremum of weighted average of (bounded) $(\sqrt{m'}/2)$-dependent random fields, where weights belong to a given class, has sub-gaussian tails. As mentioned earlier, we expect this to be useful in $m$-approximation approaches that are used for deriving limit theorems for dependent random variables and to obtain their empirical process extensions. 
Here, we use it to control the centered empirical averages $\M_n - \bar{M}_n$. The proof of the above result is outlined in Section \ref{c5pf:chainmdep} of the Appendix.


We are now in a position to deduce a bound on the rate of convergence of $d(\hat{S}_n,S_0)$. 
\begin{thm}\label{c5prp:ratemureg}
Let $\nu_n = \max\left\{h_n, (n h_n^2)^{-1/(2p)}\right\}$. For some $K_1>0$, and $\delta _n = K_1 \nu_n$,    $P^*\left( d(\hat{S}_n, S_0) >  \delta_n \right) \rightarrow 0$ as $n \rightarrow \infty$. 
\end{thm}
{\it Proof.} Let $k_n$ be the smallest integer such that $2^{k_n+1} \delta_n \geq 1$.
For $0 \leq k \leq k_n$, let $\mathcal{S}_{n,k} = \left\{S : S \in \mathcal{S},\ 2^k \delta_n < d(S, S_0 ) \leq  2^{k+1} \delta_n\right\}$.
As, $\hat{S}_n$ is the minimizer for $\M_n$, 
\begin{eqnarray*}
{ P^* \left(d(\hat{S}_n, {S}_0) > \delta_n\right) }  &\leq& \sum_{k=0}^{k_n} P^*\left( \inf_{A \in \mathcal{S}_{n,k}}\M_n(A) - \M_n({S}_0) \leq 0 \right). 
\end{eqnarray*}
The sum on the right side can be written as: 
\begin{equation} 
\sum_{k=0}^{k_n} P^*\left(\sup_{A \in \mathcal{S}_{n,k}} \left| (\M_n - M)(S) -  (\M_n - M)(S_0) \right| > \inf_{A \in \mathcal{S}_{n,k}} \left({M}(S) - {M}({S}_0) \right)  \right). 
\end{equation}
For $c(\gamma) = \min (\gamma - 1/2, 1 - \gamma)$, $ M(S)  - M(S_0) \geq c(\gamma) \lambda(S \triangle S_0)$,
and hence \eqref{c5eq:simp} is bounded by
\begin{eqnarray}
\lefteqn{\sum_{k=0}^{k_n} P^*\left(\sup_{A \in \mathcal{S}_{n,k}} \left| (\M_n - \bar{M}_n)(S) -  (\M_n - \bar{M}_n)(S_0) \right| > c(\gamma) 2^{k-1} \delta_n \right) } \nonumber \\
&& + \sum_{k=0}^{k_n} 1\left[ \sup_{A \in \mathcal{S}_{n,k}} \left| (\bar{M}_n - {M})(S) -  (\bar{M}_n - {M})(S_0) \right| \geq c(\gamma) 2^{k-1} \delta_n   \right]. 
\label{c5eq:simp}
\end{eqnarray}

We first apply Lemma \ref{lm:constbound} to the second sum in the above display. Note that
\begin{eqnarray*}
\lefteqn{\sup_{A \in \mathcal{S}_{n,k}} \left| (\bar{M}_n - {M})(S) -  (\bar{M}_n - {M})(S_0) \right|}\\
& \leq& |\tilde{\Phi}_n(0) - 1/2| 2^{k+1}\delta_n +\min(h_n, 2^{k+1}\delta_n)  \\
& & + \left|\tilde{\Phi}_n\left(\frac{\sqrt{n h_n^2} C_0 (a_n - 2L_0 h_n )^p}{2 \sqrt{1 + \Sigma^2}} \right) - 1\right| 2^{k+1}\delta_n \\
& & + \left|\tilde{\Phi}_n\left(\frac{\sqrt{n h_n^2} \delta_0}{2 \sqrt{1 + \Sigma^2}} \right) - 1\right| 2^{k+1}\delta_n + c_1 h_n \\
& \leq &   4\left[|\tilde{\Phi}_n(0) - 1/2| +  \left|\tilde{\Phi}_n\left(\frac{\sqrt{n h_n^2} \delta_0}{2 \sqrt{1 + \Sigma^2}} \right) - 1\right| \right]    2^{k-1 } \delta_n \\
& & + \left[\frac{2 c_0 a_n + c_1 h_n}{\delta_n} + 4 \left|\tilde{\Phi}_n\left(\frac{\sqrt{n h_n^2} C_0 (a_n - 2L_0 h_n )^p}{2 \sqrt{1 + \Sigma^2}} \right) - 1\right| \right] 2^{k-1} \delta_n.
\end{eqnarray*}
Fix $0 < \eta < c(\gamma)/8$. For large $n$, $|\tilde{\Phi}_n(0) - 1/2| + c_1 /(\sqrt{n}\delta_n )< \eta$. Choose $c_\eta$ such that $a_n= c_\eta \nu_n > \left[2\tilde{\Phi}_n^{-1}(1 - \eta)\sqrt{1 + \Sigma^2}/ ( C_0 \sqrt{n h_n^2}) \right]^{1/p} + (2 L_0) h_n$.  Then the coefficient of $2^{k-1} \delta_n$ in the above display is bounded by
$$ 8\eta +  \frac{2 c_0 c_\eta + c_1 }{K_1} < c(\gamma),$$
when $K_1 > (2 c_0 c_\eta + c_1 )/ (c(\gamma) - 8 \eta) $. Hence, for a suitably large choice of $K_1$ each term in the second sum of \eqref{c5eq:simp} is zero. 

We now apply Theorem \ref{thm:vdgr} to each term in the first sum of \eqref{c5eq:simp}. For this we use the following claim to obtain a bound on the entropy of the class $\mathcal{S}_{n,k}$.

{\bf Claim A.} We claim that $\sup_{S_1, S_2 \in \mathcal{S}}|\tilde{d}_n^2(S_1, S_2) - \lambda(S_1 \triangle S_2)| = O(h_n)$ and 
that $H(u, \{B \triangle S_0: B \in \mathcal{S}_{n,k}\}, \tilde{d}_n) \leq A_1 (u- c_2 h_n)^{-1}$
for constants $c_2>0$ and $A_1>0$.

We first use the above claim to prove the result. 
 As a consequence of Claim A, 
 $\sup_{A \in \{B \triangle S_0: B \in \mathcal{S}_{n,k}\}} \| A \|_{n} \leq R_{n,k} := (2^{k+1} \delta_n + c_3 h_n)^{1/2}$, for some $c_3 >0$.   Using Theorem \ref{thm:vdgr} with $\delta_1 = c(\gamma) 2^{k-1} \delta_n$, $\delta_2 = 8 c_2 h_n$, 
we arrive at the condition
 \begin{eqnarray}
\sqrt{n/m'} (c(\gamma) 2^{k-1} \delta_n - 8 c_2 h_n) \gtrsim (R_{n,k} + c_4 h_n)^{1/2} \vee R_{n,k}, \nonumber
\end{eqnarray}
for some $c_4 >0$. As $\delta_n \gtrsim \nu_n$, this translates to  $ (2^{k-1} c(\gamma) )^4 \delta_n^3  \gtrsim (2^k + c_5) h_n^4$ for some $c_5 >0$. This holds for all $k$ when $ \delta \gtrsim h_n^{4/3}$ which is true as $\delta_n \gtrsim h_n$. 
Hence, we can bound the first sum in \eqref{c5eq:simp} by 
 \begin{eqnarray}
\sum_{k=0}^{k_n} 5 C \exp \left[-\frac{n \left(c(\gamma) 2^{k-1} \delta_n - 8 c_2 h_n\right)^2}{C m' (2^{k+1} \delta_n + c_3 h_n)^{1/2}}\right].
\label{c5eq:ineqp}
\end{eqnarray}
Consequently, the display in \eqref{c5eq:ineqp} is bounded by  
\begin{eqnarray*}
\sum_{k=0}^{\infty} 5 C \exp \left[-\frac{ \left(c(\gamma)2^{k-1} - c_6 \right)^2 \delta_n }{C h_n^2 (2^{k+1} + c_7)  }\right],
\end{eqnarray*}
for some constants $c_6$, $c_7 >0$. 
As $\delta_n / h_n^2 \gtrsim h_n^{-1} \rightarrow \infty$, we get the result. 


{\it Proof of {\bf Claim A}.} Note that $\tilde{d}^2(S_1, S_2) = \tilde{d}^2(S_1 \triangle S_2, \phi) = \max_{1 \leq i, j \leq \left\lceil \sqrt{m'}\right\rceil} Q^{ij}_n (S_1  \triangle S_2)$, where  $Q^{ij}_n$ is the discrete uniform measure on the points $\{ x_{k_i l_j}: k_i = i + k \left\lceil \sqrt{m'}\right\rceil \leq m,\  l_j = j + l \left\lceil \sqrt{m'}\right\rceil\leq m \}$.   Note that each $Q^{ij}_n$ approximates Lebesgue measure at resolution of rectangles of length $m/ m' = O(h_n)$. The rectangles that intersect with the boundary of a set $S$ account for the difference $|Q^{ij}_n(S) - \lambda(S)|$. As argued in the proof of Lemma \ref{lm:constbound},  the error $\sup_{S \in \mathcal{S}}\max_{i,j} |Q^{ij}_n(S) - \lambda(S)|\leq \lambda(\{ x: \rho(x, \partial S) < O(h_n) \}$, which is $O(h_n)$ using \eqref{eq:thinbnd}.   

To see that $H(u, \{B \triangle S_0: B \in \mathcal{S}_{n,k}\}, \tilde{d}_n) \leq A_1 (u- c_2 h_n)^{-1}$, first, note that $H(u, \{B \triangle S_0: B \in \mathcal{S}_{n,k}\}, \tilde{d}_n) \leq H(u, \mathcal{S}, \tilde{d}_n)$.  
For any convex set $S$, it can be shown from arguments analogous to those in the proof for Lemma \ref{lm:constbound} that for some $c_2>0$,
$$\max_{1 \leq i, j \leq \left\lceil \sqrt{m'}\right\rceil} Q^{ij}_n (S^\delta \backslash _\delta S) \leq \lambda (S^{(\delta + c_2 h_n)} \backslash _{(\delta + c_2 h_n)}S ) \leq c_0 (\delta +c_2 h_n).$$
If $S_1, \ldots, S_r$ are the center of the Hausdorff balls with radius $\delta$ that cover $\mathcal{S}$ (see \eqref{c5eq:hdist} in the Appendix for a definition of Hausdorff distance $d_H$), then $[_\delta S_i, S_i^\delta]$, $i\leq r$ form brackets that cover $\mathcal{S}$. The sizes of these brackets are $(c_0 (\delta +c_2 h_n))^{1/2}$ in terms of the distance $\tilde{d}_n$.
Hence, 
$$H((c_0 (\delta +c_2 h_n))^{1/2}, \mathcal{S},\tilde{d}_n) \leq H_B((c_0 (\delta +c_2 h_n))^{1/2}, \mathcal{S},\tilde{d}_n) \leq H(\delta, \mathcal{S}, d_H).$$
Letting $u =     c_0 (\delta +c_2 h_n))^{1/2}$ and using the fact that $H(\delta, \mathcal{S}, d_H) \lesssim \delta^{-1/2}$
we get {\bf Claim A}. 
\qed

As was the case with Corollary \ref{cor:expbnd}, Proposition \ref{c5prp:ratemureg} extends to the following result in an identical manner.
\begin{cor}\label{cor:expbndreg} 
For the choice of $\delta_n$ given in Proposition \ref{c5prp:ratemureg}, 
\be 
\lim \sup_{n \rightarrow \infty } \sup_{\mu \in \mathcal{F}_p} E_{\mu}^* \left[\delta_n^{-1} d(\hat{S}_n,S_0)\right] < \infty. 
\ee
\end{cor}

\begin{rmk}\label{rm:bias}
The best rate at which the distance $d(\hat{S}_n,S_0)$ goes to zero corresponds to $h_n \sim (n h_n)^{-1/(2p)}$ which yields $\nu_n \sim h_n = h_0 n^{-1/{2(p+1)}}$. This is slower than the rate we deduced in the dose-response setting in terms of the total budget ($N^{-2/(4p +3)}$). The difference in the rate from the dose-response setting is accounted for by the bias in the smoothed kernel estimates. The regression setting is approximately equivalent to a dose-response model having $(2L_0 h_n)^{-2}$ (effectively) independent covariate observations and $n (2 L_0h_n)^2$ (biased) replications. These replications correspond to the number of observations used to compute $\hat{\mu}$ at a point. If we compare Lemmas \ref{lm:constboundmn} and \ref{lm:constbound}, these biased replications add an additional term of order $h_n$  which is absent in the dose-response setting. 
This puts a lower bound on the rate at which the set $S_0$ can be approximated. In contrast, the rates coincide for the dose-response and the regression settings in the one-dimensional case; see \cite{malbansen13ds} and \cite{malbansen13}. This is due to the fact that in one dimension, the bias to standard deviation ratio ($b_n/ (1/\sqrt{n\,b_n})$), where $b_n = h_n$ is the volume of the bin, is of smaller order compared to that in two dimensions ($\sqrt{b_n}/ (1/\sqrt{n b_n})$) for estimating $\hat{\mu}$ (since in the 2d case $b_n = h_n^2$, $h_n$ being the bandwidth). In a nutshell, the curse of dimensionality kicks in at dimension 2 itself in this problem.    
\end{rmk}

\subsection{Extension to the case of an unknown $\tau_0$}\label{sec:extntau2}
While we deduced our results under the assumption of a known $\tau_0$, in real applications $\tau_0$ is generally unknown. Quite a few extensions are possible in this situation. For example, in the dose-response setting, if $S_0$ can be safely assumed to contain a positive $F$-measure set $U$, then a simple averaging of the $\bar{Y}$ values realized for $X$'s in $U$ would yield a $\sqrt{mn}$-consistent estimator of $\tau_0$. If a proper choice of $U$ is not available, one can obtain an initial estimate of $\tau_0$ in the dose--response setting as
$$\hat{\tau}_{init} = \mathop{\mbox{argmin}}_{\tau \in \R} \P_n \left[\Phi\left(\sqrt{m}(\bar{Y} - \tau)\right)- \frac{1}{2} \right]^2.$$
This provides a consistent estimate of $\tau_0$ under mild assumptions. A $\sqrt{mn}$-consistent estimate of $\tau_0$ can then be found by using $\hat{\tau}_{init}$ to compute $\hat{S}_n$ and then averaging the $\bar{Y}$ value for the $X$'s realized in $ _\delta \hat{S}_n$ for a small $\delta >0$. Note that this leads to an iterative procedure where this new estimate of $\tau$ is used to update the estimate of $\hat{S}_n$.
It can be shown that the rate of convergence remains unchanged if one imputes a $\sqrt{mn}$-consistent estimate of $\tau_0$. A brief sketch of the following result is given in Section \ref{c5pf:extnuntau}. 
\begin{prop}\label{c5th:extnuntau}
Let $\hat{S}_n$ now denote the  minimizer of 
$$ \M_{n}(S, \hat{\tau}) = \P_n \left[\left\{\Phi\left({\sqrt{m}(\bar{Y} - \hat{\tau})}\right) - \gamma\right\}1_S(X)\right]\,,$$
where $\sqrt{mn}\,(\hat{\tau} - \tau_0) = O_p(1)$. For $m = m_0 n^{\beta}$ and $\delta_n$ as defined in Theorem \ref{c5prp:ratemu}, we have $P\left[ d(\hat{S}_n, S_n) > \delta_n \right] \rightarrow 0.$ 
\end{prop}
In the regression setting as well, an initial consistent estimate of $\tau_0$ can be computed as 
$$\hat{\tau}_{init} = \mathop{\mbox{argmin}}_{\tau \in \R} \frac{1}{n} \sum_{k,l}\left[\Phi\left(\sqrt{n h_n^2}(\hat{\mu}(x_{kl}) - \tau)\right)- \frac{1}{2} \right]^2$$ which can then be used to yield a $\sqrt{n}$-consistent estimate of $\tau_0$ using the iterative approach mentioned above. We have the following result for the rate of convergence of $\hat{S}_n$ in the regression setting.
\begin{prop}\label{c5th:extnuntaureg}
Let $\hat{S}_n$ now denote the  minimizer of 
$$ \M_{n}(S, \hat{\tau}) = \frac{1}{n} \sum_{k,l} \left[\left\{\Phi\left({\sqrt{n h_n^2}(\hat{\mu}(x_{kl})- \hat{\tau})}\right) - \gamma\right\}1_S(X)\right]\,,$$
where $\sqrt{n}\,(\hat{\tau} - \tau_0) = O_p(1)$. For $\delta_n$ as defined in Theorem  \ref{c5prp:ratemureg},  $P\left[ d(\hat{S}_n, S_n) > \delta_n \right] \rightarrow 0.$ 
\end{prop}
The proof is outlined in Section \ref{c5pf:extnuntaureg} of the Appendix.



\section{Discussion}\label{sec:lvl}
{\it Extensions to non-convex baseline sets.} Although we essentially address the situation where the baseline set is convex for dimension $d=2$, our approach extends past convexity and the two-dimensional setting in the presence of an efficient algorithm and for suitable collections of sets. For example, let $\tilde{\mathcal{S}}$ denote such a collection of subsets of $[0,1]^d$ sets such that 
$$\tilde{S}_n = \mathop{\mbox{argmin}}_{S \in \tilde{\mathcal{S}}} \M_n(S)$$ is easy to compute. Here, $\mu$ is a real-valued function from $[0,1]^d$ and $S_0 = \mu^{-1}(\tau_0)$ is assumed to belong to the class $\tilde{\mathcal{S}}$. Then the estimator $\tilde{S}_n$ has the following properties in the dose-response setting.
\begin{prop}
Assume that $S_0$ is the unique minimizer (up to $F$-null sets) of the population criterion function $M_F$ defined in \eqref{c5eq:constfn}. Then $d_F(\tilde{S}_n, S_0)$ converges in probability to zero. Moreover, assume that there exists a constant $\bar{c} >0$ such that  $F(S^\epsilon \backslash _\epsilon S) \leq \bar{c}\epsilon$ for any $\epsilon >0$ and $S_0 \in \tilde{\mathcal{S}}$, and 
$$ H_B(u, \tilde{\mathcal{S}}, L_2(P)) \lesssim u^{-r} \mbox{ for some $r <2$} .$$
Then, $P\left(d_F(\tilde{S}_n, S_0) > \tilde{\delta}_n\right)$ converges to zero where $\tilde{\delta}_n = K_1 \max ( n^{-2/(2+r)}, m^{-1/(2p)})$ for some $K_1 >0$.
\end{prop}
The proof follows along lines identical to that for Theorem \ref{c5prp:ratemu}. Note that the condition $F(S^\epsilon \backslash _\epsilon S) \leq \bar{c} \epsilon$  was needed to derive Lemma \ref{lm:constboundmn}. This assumption simply rules out sets with highly irregular or non-rectifiable boundaries. Also, the dependence of the rate on the dimension typically comes through $r$ which usually grows with $d$. A similar result can be established in the regression setting as well.

{\it Connection with level-set approaches.} Note that minimizing  $\mathbb{M}_{n}(S)$ in the dose-response setting is equivalent to minimizing
\begin{eqnarray*}
 \tilde{\mathbb{M}}_{ n}(S) & = &  \mathbb{M}_{ n}(S)  - \frac{1}{2}\sum_{i=1}^N \left( \frac{1}{4}  - p_{m,n}(X_i) \right) \\
 & = & \sum_{i=1}^n \frac{{1}/{4}  - p_{m,n}(X_i)}{2} \left[1( X_i \in S) - 1( X_i \in S^c) \right]\\
\end{eqnarray*}
This form is similar to an empirical risk criterion function that is used in \citet[equation (7)]{WN07} in the context of a level-set estimation procedure. It can be deduced that our baseline estimation approach ends up finding the level set $ S_m = \{ x: E\left[ p_{m,n} (x)\right] > 1/4 \}$ from i.i.d. data $\{ p_{m,n}(X_i), X_i \}_{i=1}^n$ with $0 \leq p_{m,n}(X_i) \leq 1$. As $m \rightarrow \infty$, $S_m$'s decrease to $S_0$, which is the target set. Hence, any level-set approach could be applied to transformed data $\{ p_{m,n}(X_i), X_i \}_{i=1}^n$ to yield an estimate for $S_m$ which would be consistent for $S_0$.   
Moreover, a similar connection between the two approaches can be made for the regression setting, however the i.i.d. flavor of the observations present in the dose-response setting is lost as $\{ p_{n}(x_{kl})\}_{1 \leq k,l \leq  m}$ are dependent. While the algorithm from \cite{WN07} can be implemented to construct the baseline set estimate, it is far from clear how the theoretical properties would then translate to our setting given the dependence of the target function $E \left[p_{m,n} (x)\right]$ on $m$ in the dose-response setting and the dependent nature of the transformed data in the regression setting. 

In \cite{SD07}, the approach to the level set estimation problem, using the criterion in \cite{WN07}, is shown to be equivalent to a {\it cost-sensitive classification} problem. This problem involves random variables $(X, Y, C) \in \mathbb{R}^d \times \{0, 1\} \times \mathbb{R}$, where $X$ is a feature, $Y$ a class and $C$ is the cost for misclassifying $X$ when the true label is $Y$. Cost sensitive classification seeks to minimize the expected cost
\be 
R(G) = E(C\ 1(G(X) \neq  Y)),
\label{c5eq:closs}\ee
where $G$, with a little abuse of notation, refers both to a subset of $\mathbb{R}^d$ and $G(x) = 1( x \in G)$. With $C = |\gamma - Y|$ and $\tilde{Y} = 1(Y \geq \gamma)$, the objective of the cost-sensitive classification, based on $(X, \tilde{Y}, C)$, can be shown to be equivalent 
to minimizing the excess risk criterion in \cite{WN07}. So, approaches like support vector machines (SVM) and $k$-nearest neighbors ($k$-NN), which can be tailored to solve the cost-sensitive classification problem (see \cite{SD07}), are relevant to estimating level sets, and thus provide alternative ways to solve the baseline set estimation problem. Since the {\it loss function} in \eqref{c5eq:closs} is not smooth, one might prefer to work with surrogates. Some results in this direction can be found in \cite{S11}. 

{\it Adaptivity.} We have assumed knowledge of the order
of the regularity  $p$ of $\mu$ at $\partial S_0$, which is required to achieve the optimal rate
of convergence, though not for consistency. The knowledge of $p$ dictates the allocation between $m$ and $n$ in the dose-response setting and the choice of the bandwidth $h_n$ in the regression setting for attaining the best possible rates. 
When $p$ is unknown, the adaptive properties of dyadic trees (see \cite{WN07} and \cite{SNC09}) could conceivably be utilized to develop a near-optimal approach. However, this is a hard open problem and will be a topic of future research.

\appendix

\section{Proofs}
\subsection{Proof of Theorem \ref{c5th:constmn}}\label{c5pf:constmn}
Here, we establish consistency with respect to the (stronger) Hausdorff metric,
\be\label{c5eq:hdist}
 d_H(S_1,S_2) = \max \left[ \sup_{x \in S_1} \rho( x, S_2), \sup_{x \in S_2} \rho( x, S_1)\right].
\ee
Moreover, we would only require $\min(m,n) \rightarrow \infty$ instead of taking $m$  to be of the form $m_0 n^\beta$, $\beta >0$.

To exhibit the dependence on $m$, we will denote $\M_n$ by $\M_{m,n}$. 
Recall that $M_m(S) = E\left[\M_{m,n}(S)\right]$ converges to $M(S)$ for each $S \in \mathcal{S}$. 
Also, $\mbox{Var} (\M_{m,n}(S)) = (1/n) \mbox{Var} \left((\Phi(\sqrt{m} \bar{Y_1}) - \gamma) 1_{S}(X)\right) \leq 1/n$ which converges to zero. Hence, $\M_{m,n}(S)$ converges in probability to $M(S)$ for any $S \in \mathcal{S}$, as $\min(m, n) \rightarrow \infty$. 

The space $(\mathcal{S}, d_H)$ is compact (Blaschke Selection theorem) and $M$ is a continuous function on $\mathcal{S}$. The desired result will be a consequence of argmin continuous mapping theorem \citep[Theorem 3.2.2]{VW96} provided we can justify that $\sup_{S \in \mathcal{S}} | \M_{m,n}(S) - M (S)| $ converges in probability to zero. To this end, let $$\M_{m,n}^1(S) = \M_{m,n}(S) + \P_n \gamma 1_X(S) = \P_n \Phi\left(\sqrt{m}\bar{Y}\right)  1_S(X) $$ and $\M^1(S) = \M(S) + P \gamma 1_X(S)$. 
Note that 
$$\sup_{S \in \mathcal{S}} | \M_{m,n}(S) - M (S)|  \leq \gamma \sup_{S \in \mathcal{S}} |(\P_n - P)(S)| + \sup_{S \in \mathcal{S}} | \M^1_{m,n}(S) - M^1 (S)| .$$
The first term in the above expression converges in probability to zero \citep{R62}. As for the second term, note that $\M^1_{m,n}(S)$ converges in probability to  $M^1 (S)$ for each $S$ and $\M^1_{m,n}$ is monotone in $S$, i.e., $\M^1_{m,n} (S_1) \leq \M^1_{m,n} (S_2)$ whenever $S_1 \subset S_2$. As the space $(\mathcal{S}, d_H)$ is compact, there exist $S(1), \ldots, S({l(\delta)})$ such that $\sup_{S \in \mathcal{S}} \min_{1\leq l \leq l(\delta)} d_H(S,S(l)) < \delta$,  for  any  $\delta >0$. Hence,
\begin{eqnarray*}
\lefteqn{\sup_{S \in \mathcal{S}} | \M^1_{m,n}(S) - M^1 (S)| }\\
& = & \max_{1\leq l \leq l(\delta)} \sup_{d_H(S, S(l)) < \delta} | \M^1_{m,n}(S) - M^1 (S)| \\
& \leq & 2 \max_{1\leq l \leq l(\delta)} \sup_{d_H(S, S(l)) < \delta} | \M^1_{m,n}(S) - M^1 (S(l))| \\
& \leq & 2 \max_{1\leq l \leq l(\delta)} \max(| \M^1_{m,n}(_\delta{(S(l))}) - M^1 (S(l))|, | \M^1_{m,n}({(S(l))}^\delta) - M^1 (S(l))|).  
\end{eqnarray*}
 The right side in the above display converges in probability to $2 \max_{1\leq l \leq l(\delta)} [\max(| M^1(_\delta (S(l))) - M^1 (S(l))|, | M^1({(S(l))}^\delta) - M^1 (S(l))|)]$ can be made arbitrarily small  by choosing small $\delta$ (as $M^1$ is continuous).
Also, as the map $S \mapsto d_F(S, S_0)$ from $(\mathcal{S}, d_H)$ to $\R$ is continuous, we have consistency in the $d_F$ metric as well. This completes the proof. \qed

%
\subsection{Proof of Theorem \ref{c5th:constreg}} \label{c5pf:constreg}
In light of what has been derived in the proof of Theorem \ref{c5th:constmn}, it suffices to show that 
$\M_n(S)$ converges in probability to $M(S)$. 
Note that, 
$$ E \left[\Phi \left(\sqrt{n h_n^2} \hat{\mu}(x_{kl})\right) \right]= \tilde{\Phi}_n \left(\frac{\sqrt{n h_n^2}\bar{\mu}(x_{kl}) }{\sqrt{1 + \Sigma^2_n(x_{kl})}}\right) .$$
For $x \in  \{ (x_1, x_2): k/m  \leq x_1 < (k+1)/m,   l/m  \leq x_2 < (l+1)/m \}$, let 
$$\hat{f}_n (x) = \widetilde{W}_{kl} \mbox{ and } f_n(x) = E \left[\hat{f}_n (x) \right] =  \tilde{\Phi}_n \left(\frac{\sqrt{n h_n^2}\bar{\mu}(x_{kl}) }{\sqrt{1 + \Sigma^2_n(x_{kl})}}\right) - \gamma.$$   Then $\bar{M}_n (S) = \int_S f_n(x) dx$. For any fixed $x$ in the  interior of the set $S_0$, $f_n(x) = \tilde{\Phi}_n(0)$ for sufficiently large $n$ which converges to $1/2$. As $\mu$ is continuous, for any fixed $x \notin S_0$, $\mu_{x,\delta_x} = \inf\{ \mu(y): \rho(x,y) < \delta_x \}  >0$ for some $\delta_x >0$. Hence  $f_n(x) \geq \Phi(\sqrt{n h_n^2} \mu_{x,\delta_x})$ converges to 1.  
Also, $|f_n(x)| \leq 1$ and hence,  $\bar{M}_n(S)$ converges to $M(S)$ by the Dominated convergence theorem.

Moreover, 
$$\mbox{Var}(\M_n(S)) \leq \frac{1}{n^2} \sum_{k,l,k',l'}\mbox{ Cov} \left(\hat{f}_n(x_{k,l}), \hat{f}_n(x_{k',l'})\right) .$$
As $|\hat{f}_n(x_{k,l})| \leq 1$, and $\hat{f}_n(x_{k,l})$ and $\hat{f}_n(x_{k',l'})$ are independent whenever $\min\{|k -k'|, |l - l'| \} > 2 L_0 m h_n$, we have 
$$ \sum_{k',l'}\mbox{ Cov} \left(\hat{f}_n(x_{k,l}), \hat{f}_n(x_{k',l'})\right) \lesssim (m h_n)^2 = n h_n^2,$$
for any fixed $k$ and $l$. Hence, $\mbox{Var}(\M_n(S)) $is bounded (up to a constant) by $n (n h_n ^2)/n^2$ which converges to zero. Hence, $\M_n(S)$ converges in probability to $M(S)$, which completes the proof.
\qed

\subsection{Proof of Lemma \ref{lm:constbound}}\label{cbapp}  Let  $Bin_{kl} = \{ x = (x_1, x_2):\ k/m  \leq x_1 < (k+1)/m,   l/m  \leq x_2 < (l+1)/m \}$. 
Recall that
\begin{eqnarray*}
\lefteqn{{M}(S) - {M} (S_0) }\\
& = & \int \left\{(1/2) 1_{S_0}(x) + 1_{S_0^c}(x)   - \gamma \right\} \left\{1_{S}(x) - 1_{S_0}(x) \right\} dx\\
& = & \sum_{0 \leq k,l \leq (m-1)} \int_{Bin_{kl}} \left\{(1/2) 1_{S_0}(x) + 1_{S_0^c}(x)   - \gamma\right\} \left\{1_{S}(x) - 1_{S_0}(x) \right\} dx\\
& = & \sum_{0\leq k,l \leq (m-1)} \left[\int_{Bin_{kl} \cap (S_0\cap S^c)} \left\{\frac{1}{2}- \gamma\right\}  dx + \int_{Bin_{kl} \cap (S_0^c\cap S)} \left\{1- \gamma\right\}  dx \right]\\ 
& = & \sum_{k,l:x_{kl} \in \mathcal{I}_n } \left[\int_{Bin_{kl} \cap (S_0\cap S^c)} \left\{\frac{1}{2}- \gamma\right\}   dx + \int_{Bin_{kl} \cap (S_0^c\cap S)} \left\{1- \gamma\right\}  dx \right] + e^{(1)}_n.
\end{eqnarray*}
Here, $e^{(1)}_n$ is the remainder term arising out of replacing the sum of all choices of $k$ and $l$ to sum over $\{(k,l): x_{kl} \in \mathcal{I}_n\}$. As the integrands in the above sum are bounded by 1, $|e^{(1)}_n| \leq  \lambda([0,1]^2 \backslash \mathcal{I}_n) = O(h_n)$. Also, 
\begin{eqnarray*}
\lefteqn{\bar{M}_n(S) - \bar{M}_n (S_0) } \\
& = &\frac{1}{n}\sum_{k,l}\left\{\tilde{\Phi}_n\left(\frac{\sqrt{n h_n^2} \bar{\mu}(x_{kl})}{\sqrt{1 + \Sigma^2_n(x_{kl})}} \right) - \gamma\right\} \left\{1_{S}(x_{kl}) - 1_{S_0}(x_{kl}) \right\} \\
& = & \sum_{k,l}\int_{x \in Bin_{kl}} \left\{\tilde{\Phi}_n\left(\frac{\sqrt{n h_n^2} \bar{\mu}(x_{kl})}{\sqrt{1 + \Sigma^2_n(x_{kl})}} \right) - \gamma\right\} \left\{1_{S}(x_{kl}) - 1_{S_0}(x_{kl}) \right\} dx.
\end{eqnarray*}
Consequently,
{\allowdisplaybreaks \begin{eqnarray}
\lefteqn{(\bar{M}_n - {M})(S) -  (\bar{M}_n - {M})(S_0) } \hspace{1in}\nonumber\\
& =& \sum_{k,l} \int_{x \in Bin_{kl} \cap (S_0 \cap S^c) } \left(\tilde{\Phi}_n\left(\frac{\sqrt{n h_n^2} \bar{\mu}(x_{kl})}{\sqrt{1 + \Sigma^2_n(x_{kl})}} \right) - \frac{1}{2} \right)dx   \nonumber \\
& & + \sum_{k,l} \int_{x \in Bin_{kl} \cap (S_0^c \cap S) } \left(\tilde{\Phi}_n\left(\frac{\sqrt{n h_n^2} \bar{\mu}(x_{kl})}{\sqrt{1 + \Sigma^2_n(x_{kl})}} \right) - 1 \right)dx  \nonumber \\
& & + e^{(2)}_n,
\label{c5eq:frstbnd}
\end{eqnarray}
where $e^{(2)}_n$, the contribution of the terms at $\partial(S_0 \cap S)$ along with $e^{(1)}_n$, is bounded by
 \begin{eqnarray*}
|e^{(1)}_n| +  \sum_{k,l: Bin_{kl} \cap \partial(S_0 \cap S) \neq \phi} \int_{Bin_{kl} } 2 dx.
\end{eqnarray*}
This is further bounded by $2 \lambda (\{ x : \rho(x, \partial S_0 ) < 2/m \} ) + 2 \lambda (\{ x : \rho(x, \partial S ) < 2/m \} )$ which is at most $2 c_0 (2/m) + 2 c_0 (2/m) = 8 c_0/ m$ using \eqref{eq:thinbnd} ($\lambda\{ x : \rho(x, \partial S ) < \alpha \} \leq \lambda(S^\alpha \backslash _\alpha S)$ for any $\alpha >0$). Hence, for some $\tilde{c}_1
>0$, 
$$ |e^{(2)}_n| \leq O(h_n) + 8 c_0/m \leq \tilde{c_1} h_n.$$ This contribution is accounted for in the last term of \eqref{c5eq:meandiff}.

We now study the contribution of  the other terms in the right side of \eqref{c5eq:frstbnd}.  
Note that the integrand in the first sum in the right side of \eqref{c5eq:frstbnd} is precisely $(\tilde{\Phi}_n(0) - 1/2)$ whenever $Bin_{kl} \subset  _{(L_0 h_n)}S_0$ as $\bar{\mu}(x_{kl})$ is zero.  As the integrand is also bounded by 1, the first sum in the right side of \eqref{c5eq:frstbnd} is then bounded by $$|\tilde{\Phi}_n(0) - 1/2| \delta +\lambda((S_0 \backslash _{(L_0 h_n)}S_0)\cap S) \leq |\tilde{\Phi}_n(0) - 1/2| \delta +\min (c_0 L_0 h_n, \delta).$$  Choosing $c_1 = \tilde{c}_1 + c_0 L_0$, the second term on the right side of the above display is also accounted for in the last term in \eqref{c5eq:meandiff}. Further, let $S_n = \{ x: \rho(x, S_0) > a_n \}$. Note that $\lambda(S_n^c \backslash S_0) \leq c_0 a_n$ using \eqref{eq:thinbnd}. Hence, the second sum in \eqref{c5eq:frstbnd} is bounded by
\begin{eqnarray*}
\lefteqn{\int_{S_n^c \backslash S_0} 1 dx + \sum_{k,l} \int_{x \in Bin_{kl} \cap (S_n^c \cap S) } \left|\tilde{\Phi}_n\left(\frac{\sqrt{n h_n^2} \bar{\mu}(x_{kl})}{\sqrt{1 + \Sigma^2_n(x_{kl})}} \right) - 1 \right|dx  }\\
& \leq & \min(c_0 a_n, \delta) + \sum_{k,l}  \int_{x \in Bin_{kl} \cap (S_n \cap S) } \left|\tilde{\Phi}_n\left(\frac{\sqrt{n h_n^2} \bar{\mu}(x_{kl})}{\sqrt{1 + \Sigma^2_n(x_{kl})}} \right) - 1\right| dx, 
\end{eqnarray*}
To bound the second term in right side of the above, note that as $x_{kl} \in \mathcal{I}_n$,
$$\bar{\mu} (x_{kl}) = E \left[\mu (x_{kl} + h_n Z_n)\right] \left\{\frac{1}{n h_n^2} \sum_{r,s: |r|,|s| \leq L_0 m h_n }K\left(\left(\frac{r}{m}, \frac{s}{m}  \right)\right)\right\},$$
where $Z_n$ is a discrete random variable supported on $\left\{(r/m,s/m): |r|,|s| \leq L_0 m h_n \right\}$ with mass function 
$P\left[Z_n = (r/m,s/m)\right] \propto K\left(\left({r}/{m}, {s}/{m} \right)\right)$. Hence, the argument of $\tilde{\Phi}_n$ can be written as 
$$\sqrt{nh_n^2}E \left[\mu (x_{kl} + h_n Z_n)\right] \frac{ \sum_{r,s: |r|,|s| \leq L_0 m h_n }K\left(\left({r}/{m}, {s}/{m} \right)\right)}{n h_n^2 \sqrt{1 + \tilde{\Sigma}^2_n}}$$ 
%
%
Note that 
$$\frac{ \sum_{r,s: |r|,|s| \leq L_0 m h_n }K\left(\left({r}/{m}, {s}/{m} \right)\right)}{n h_n^2 \sqrt{1 + \tilde{\Sigma}^2_n}} = \frac{1}{\sqrt{1 + \Sigma^2}} + o(1), $$
uniformly in $k$ and $l$ for $x_{kl} \in S \cup S_0$.
For $x_{kl} \in S_n \cap S_0$ and $a_n < \kappa_0$, when $\rho(x_{kl}+ h_n Z_n, S_0) < \kappa_0$, by triangle inequality, 
$$\mu ( x_{kl} + h_n Z_n) \geq C_0 \rho(x_{kl} + h_n Z_n)^p \geq C_0  (\rho( x_{kl}, S_0) -  \rho( x_{kl}, x_{kl} + h_n Z_n))^p.$$
As $\rho( x_{kl}, x_{kl} + h_n Z_n) \leq 2 L_0 h_n$, 
$$\mu ( x_{kl} + h_n Z_n) > C_0 (a_n - 2L_0 h_n )^p.$$ 
On the other hand, when $\rho(x_{kl}+ h_n Z_n, S_0) \geq \kappa_0$,
$\mu ( x_{kl} + h_n Z_n) > \delta_0$. Consequently, for $x_{kl} \in S_n \cap S_0$, we get 
\begin{eqnarray*}
{\left|\tilde{\Phi}_n\left(\frac{\sqrt{n h_n^2} \bar{\mu}(x_{kl})}{\sqrt{1 + \Sigma^2_n(x_{kl})}} \right) - 1\right|  }
&\leq& \left|\tilde{\Phi}_n\left(\frac{\sqrt{n h_n^2} C_0 (a_n - 2L_0 h_n )^p}{2 \sqrt{1 + \Sigma^2}} \right) - 1\right|\\
& & + \left|\tilde{\Phi}_n\left(\frac{\sqrt{n h_n^2} \delta_0}{2 \sqrt{1 + \Sigma^2}} \right) - 1\right|.
\end{eqnarray*}
As $\lambda(S_n \cap S) < \delta$, we get the result.
\qed

\subsection{Proof of Lemma \ref{lm:onedubineq}}\label{c5pf:onedubineq}
The sum $\sum_{k,l} W_{kl} 1_S(x_{kl})$ can be written as $\sum_{1 \leq i, j \leq \left\lceil \sqrt{m'}\right\rceil} R_{i,j}$ where for $k_i$s and $l_j$s defined as in \eqref{c5eq:indices},  
each block
\be \label{c5eq:blocksum}
R_{i,j} = \sum_{\substack{k,l : 1 \leq k_i, l_j \leq m }} W_{k_i l_j} 1_S(x_{k_i l_j}) 
\ee
is a sum of $r_{ij}$ many independent random variables with 
$$ \left\lfloor m/ \left\lceil \sqrt{m'}\right\rceil \right\rfloor^2 \leq r_{ij} \leq \left\lceil m/ \left\lceil \sqrt{m'}\right\rceil \right\rceil^2 .$$
As $m / \sqrt{m'} \rightarrow \infty$,  
\be\label{c5eq:floorceil}
 n/(2m') \leq r_{ij} \leq  2n/m',
\ee
for large $n$, a fact we use frequently in the proofs. 
Note that $\sum_{1 \leq i, j \leq \left\lceil \sqrt{m'}\right\rceil} r_{ij} =n$ and hence, by convexity of $\exp(\cdot)$,
$$\exp \left(\frac{1}{n}\sum_{1\leq k,l\leq m} W_{kl} 1_S(x_{kl})\right) \leq \sum_{1 \leq i, j \leq \left\lceil \sqrt{m'}\right\rceil} \frac{r_{ij}}{n} \exp\left(\frac{R_{ij}}{r_{ij}}\right).$$
As  $| {W}_{kl} 1_{S}(x_{kl}) | \leq 1_{S}(x_{kl}) $, 
\begin{eqnarray*}
{ P \left( \frac{1}{n} \sum_{k,l} {W}_{kl} 1_{S}(x_{kl}) \geq a \right) }& \leq &  \sum_{1 \leq i, j \leq \left\lceil \sqrt{m'}\right\rceil} \frac{r_{ij}}{n} E\ \exp\left(\frac{\lambda R_{ij}}{r_{ij}} - \lambda a\right) \mbox{ and }\\
E\ \exp\left(\frac{\lambda R_{ij}}{r_{ij}} \right) & \leq & \exp\left(\frac{\lambda^2 }{8 r^2_{ij}} \sum_{\substack{k,l : 1 \leq k_i, l_j \leq m }} \left(1_S(x_{k_i l_j})\right)^2 \right).  
\end{eqnarray*}
The second bound in the above display is simply the one used in proving  Hoeffding's inequality for independent sequences 
\cite[equation (4.16)]{H63}. Consequently,
\begin{eqnarray*}
\lefteqn{ P \left( \frac{1}{n}| \sum_{k,l} {W}_{kl} 1_{S}(x_{kl}) \geq a \right) } \\
& \leq &  e^{-\lambda a} \exp\left(\frac{\lambda^2 }{8 (n/2m')^2} \max_{1 \leq i, j \leq \left\lceil \sqrt{m'}\right\rceil} \left[\sum_{\substack{k,l : 1 \leq k_i, l_j \leq m }} 1_S(x_{k_i l_j}) \right]\right)
\end{eqnarray*}
Picking 
$$\lambda = \frac{a  (n/m')}{ \max_{1 \leq i, j \leq \left\lceil \sqrt{m'}\right\rceil} \left[(1/r_{ij})\sum_{\substack{k,l : 1 \leq k_i, l_j \leq m }} 1_S(x_{k_i l_j})\right]} $$
and paralleling the above steps to bound $P \left(({1}/{n}) \sum_{k,l} {W}_{kl} 1_{S}(x_{kl}) \leq a \right) $, 
we get 
\begin{eqnarray*}
\lefteqn{ P \left( \frac{1}{n}\left| \sum_{k,l} {W}_{kl} 1_{S}(x_{kl}) \right| \geq a \right) }\hspace{0.2in}\\
&\leq & 2 \exp \left[-\frac{n a^2}{16 m' \max_{1 \leq i, j \leq \left\lceil \sqrt{m'}\right\rceil} \left[(1/r_{ij})\sum_{\substack{k,l : 1 \leq k_i, l_j \leq m }} 1_S(x_{k_i l_j}) \right]}\right]. 
\end{eqnarray*}
Using the definition of $\tilde{d}_n$, the result follows.

\subsection{ Proof of Theorem \ref{thm:vdgr}} \label{c5pf:chainmdep}
%
Let $ S = \min\{ s \geq 1: 2^{-s}R \leq \delta_2/2 \}$. By means of condition \eqref{c5eq:conditionch}, we can choose $C$ to be a constant large enough so  that
\be
\sqrt{n/m'}(\delta_1 - \delta_2) \geq 48 \sum_{s=1}^S 2^{-s} R H^{1/2} (2^{-s} R, \mathcal{G}, \tilde{d}_n) \vee (1152 \log 2)^{1/2} (4 m' )R. \nonumber
\ee
We denote the class of functions ${\mathcal{G}}$ by $\{ g_\theta: \theta \in \Theta \}$ for convenience.
Let $\{ g_j^s\}_{j=1}^{N_s}$ be a minimal $2^{-s} R$-covering set of $\tilde{\mathcal{S}}$, $s = 0, 1, \ldots$. So, $N_s = N(2^{-s}R, \tilde{\mathcal{S}}, \tilde{d}_n)$. For, any $\theta \in \Theta$, let $g_\theta^s$ denote approximation of $g_\theta$ from the collection $\{ g_j^s\}_{j=1}^{N_s}$. As $|W_{kl}| \leq 1$, applying Cauchy-Schwartz to each block $R_{i,j}$ defined in \eqref{c5eq:blocksum}  and using \eqref{c5eq:floorceil} yields
\begin{eqnarray*}
\left| \frac{1}{n} \sum_{k,l} V_{kl} (g_\theta (x_{kl}) - g_\theta^S(x_{kl})) \right| \leq  2 \tilde{d}_n(g_\theta, g_\theta^S) \leq \delta_2.
\end{eqnarray*}
Hence, it suffices to prove the exponential inequality for 
$$ P\left(\max_{j= 1,\ldots, N_s} \left| \frac{1}{n} \sum_{k,l} V_{kl} g_j^S(x_{kl}) \right|  \geq \delta_1 - \delta_2 \right).$$
Next, we use a chaining argument. Note that $g_\theta^S = \sum_{s=1}^S (g_\theta^s - g_\theta^{s-1} )$. By triangle inequality,
$$ \tilde{d}_n (g_\theta^s, g_\theta^{s-1}) \leq \tilde{d}_n (g_\theta^s, g_\theta) + \tilde{d}_n (g_\theta, g_\theta^{s-1})  \leq 3 (2^{-s} R).$$
Let $\eta_s$ be positive numbers satisfying $\sum_{s\leq S} \eta_s \leq 1$. Then,
\begin{eqnarray}
\lefteqn{P^*\left(\sup_{\theta \in \Theta} \left| \frac{1}{n} \sum_{s=1}^S \sum_{k,l} V_{kl} (g_\theta^s (x_{kl}) - g_\theta^{s-1}(x_{kl}))\right| \geq \delta_1 - \delta_2 \right)} \nonumber\\
& \leq & \sum_{s=1}^S P^*\left(\sup_{\theta \in \Theta} \left| \frac{1}{n} \sum_{k,l} V_{kl} (g_\theta^s (x_{kl}) - g_\theta^{s-1}(x_{kl}))\right| \geq (\delta_1 - \delta_2)\eta_s \right) \nonumber\\
& \leq & \sum_{s=1}^S 2 \exp \left[ 2 H(2^{-s}R, \tilde{\mathcal{S}}, \tilde{d}_n ) - \frac{n(\delta_1 - \delta_2)^2 \eta^2_n}{9 (16 m') 2^{-2s}R^2}\right].
\label{c5eq:bounsumprob}
\end{eqnarray}
We choose $\eta_s$ to be 
$$ \eta_s = \frac{6 \sqrt{16m'} 2^{-s}R H^{1/2}(2^{-s}R, \tilde{\mathcal{S}}, \tilde{d}_n)}{\sqrt{n}(\delta_1-\delta_2)} \vee \frac{2^{-s}\sqrt{s}}{8}.$$
 The rest of the argument is identical to that Lemma 3.2 of \cite{V00}. It can be shown that $\sum_{s \leq S} \eta_s  \leq 1$.
Moreover, the above choice of $\eta_s$ guarantees $$H(2^{-s}R, \tilde{\mathcal{S}}, \tilde{d}_n) \leq \frac{n (\delta_1 - \delta_2)^2 \eta^2_s}{36 (16 m'^2) 2^{-2s}R^2}.$$ Hence the bound in \eqref{c5eq:bounsumprob} is at most
$$ \sum_{s=1}^s  2 \exp \left[ - \frac{n(\delta_1 - \delta_2)^2 \eta^2_s}{18 (16 m') 2^{-2s}R^2}\right].$$
Next, using $ \eta_s \leq 2^{-s} \sqrt{s}/8$ and that $n(\delta_1 - \delta_2^2)/(1152 (16 m')R^2) \geq \log (2)$, it can be shown that the above display is bounded above by
\begin{eqnarray*}
\lefteqn{\sum_{s=1}^\infty 2 \exp\left[- \frac{n(\delta_1 - \delta_2)^2 s}{1152 (16 m') R^2} \right] }\\
& \leq & 2 \left(1 - \exp\left[- \frac{n(\delta_1 - \delta_2)^2 }{1152 (16 m') R^2} \right] \right)^{-1} \exp\left[- \frac{n(\delta_1 - \delta_2)^2 s}{1152 (16 m') R^2} \right]\\
& \leq & 4  \exp\left[- \frac{n(\delta_1 - \delta_2)^2 s}{1152 (16 m') R^2} \right].
\end{eqnarray*}
This completes the proof.

\subsection{Proof of Proposition \ref{c5th:extnuntau}} \label{c5pf:extnuntau}
Note that $\sqrt{mn}(\hat{\tau} - \tau_0) = O_P(1)$. So, given $\alpha > 0$, there exists $L_\alpha >0$ such that for $V_{n ,\alpha} = [\tau_0 - L_\alpha/\sqrt{mn}, \tau_0 + L_\alpha/\sqrt{mn}]$, $P[\hat{\tau} \in V_{n,\alpha}] > 1 - \alpha$. Let $\hat{S}_n (\tau)$ denote the estimate  of $S_0$ based on $\M_n(S, \tau)$. 
Then,
\begin{eqnarray*}
{ P^* \left[d(\hat{S}_n(\hat{\tau}), {S}_0) > \delta_n\right] }  & \leq & P^* \left[ d(\hat{S}_n(\hat{\tau}), {S}_0) > \delta_n , \hat{\tau} \in V_{n,\alpha}\right]  + \alpha.
\end{eqnarray*}
Following the arguments for the proof of Proposition \ref{c5prp:ratemu}, 
the outer probability on the right side can be bounded by
\begin{eqnarray*}
\sum_{k=0}^{k_n} P^*\left( \inf_{A \in \mathcal{S}_{n,k} } \M_n(A, \hat{\tau}) - \M_n({S}_0, \hat{\tau}) \leq 0, \hat{\tau} \in V_{n,\alpha} \right) 
\end{eqnarray*}
The is further bounded by: 
{\small \begin{equation} 
\sum_{k=0}^{k_n} P^*\left(\sup_{\substack{A \in \mathcal{S}_{n,k},\\ \tau  \in V_{n,\alpha}}} \left| (\M_n(S,\tau) - M(S)) -  (\M_n(S_0,\tau) - M(S_0)) \right| > \inf_{A \in \mathcal{S}_{n,k}} \left({M}(S) - {M}({S}_0) \right)  \right). 
\label{c5eq:shellmn_et}
\end{equation}}
As before, $ M(S)  - M(S_0) \geq c(\gamma) F(S \triangle S_0)$,
and hence, \eqref{c5eq:shellmn_et} is bounded by
{\small \begin{equation}
\begin{split}
\lefteqn{\sum_{k=0}^{k_n} P^*\left(\sup_{\substack{A \in \mathcal{S}_{n,k},\\ \tau  \in V_{n,\alpha}}} \left| (\M_n - {M}_m)(S,\tau) -  (\M_n - {M}_m)(S_0,\tau) \right| > c(\gamma) 2^{k} \delta_n/3 \right) } \\
&+  \sum_{k=0}^{k_n} 1\left[ \sup_{\substack{A \in \mathcal{S}_{n,k},\\ \tau  \in V_{n,\alpha}}}\left| ({M}_m (S,\tau) - {M}_m (S,{\tau_0}) ) -  ({M}_m (S_0,\tau) - {M}_m (S_0,{\tau_0}) ) \right| \geq c(\gamma) 2^{k} \delta_n/3   \right] \\
&+  \sum_{k=0}^{k_n} 1\left[ \sup_{A \in \mathcal{S}_{n,k}} \left| ({M}_m - {M})(S, \tau_0) -  ({M}_m - {M})(S_0,\tau_0) \right| \geq c(\gamma) 2^{k} \delta_n/3   \right]. 
\end{split}
\label{c5eq:simpmn_et}
\end{equation}}
The third term can be shown to be zero for sufficiently large $n$ in the same manner as in the proof of Proposition \ref{c5prp:ratemu}. 
Note that the first term can be written as 
\be
\sum_{k=0}^{k_n} P^*\left(\sup_{A \in \mathcal{S}_{n,k}, \tau \in V_{n,\alpha}} \left| \G_n g_{n,\tau}(\bar{Y}) 1_{A \triangle S_0} (X) \right| > c(\gamma) {2^{k-1} \delta_n \sqrt{n}/3 }\right),
\label{c5eq:sumineqmn_un}
\ee
where  
 $g_{n,\tau} (y) = \Phi\left(\sqrt{m} (y - \tau) \right) - \gamma$. We are now in a position to apply Theorem \ref{thm:vdgrbrk} to each term of \eqref{c5eq:sumineqmn_un}. In the setup of Theorem \ref{thm:vdgrbrk}, $N = 2^{k-1} \delta_n \sqrt{n} $ and the concerned class of functions is $\mathcal{G}_{n,k} = \{ g_{n, \tau} (\bar{Y}) 1_B(X) : B = A \triangle S_0, A \in \mathcal{S}_{n,k}, \tau \in V_{n,\alpha} \}$. For $B \in \{A \triangle S_0: A \in \mathcal{S}_{n,k}\}$, $\| g_{n,\tau} 1_B \|_{L_2(P)} \leq [E 1_B(X)]^{1/2} \leq  (2^{k+1} \delta_n)^{1/2}$. So, we can choose $R = R_{n,k} =  (2^{k+1} \delta_n)^{1/2}$. Also,
$$ H_B (u,  \{A \triangle S_0: A \in \mathcal{S}_{n,k}\}, L_2(P)) \leq A_0 u^{-1}, $$
for some constant $A_0>0$. 
 To bound the entropy of the class of functions $\mathcal{T}_n =\{g_{n, \tau} (\cdot) : \tau \in V_{n,\alpha}  \}$, 
let $\tau_0-L_\alpha/\sqrt{mn} = t_0 < t_1< \ldots < t_{r_n} = \tau_0+L_\alpha/\sqrt{mn}$ be such that $|t_i - t_{i-1}| \leq u/\sqrt{m}$, for $u>0$. Note that $r_n \leq 4 L_{\alpha} u/\sqrt{n}$.
As $\Phi$ is Lipschitz continuous of order 1 (with Lipschitz constant bounded by 1),
$$ |g_{n,\tau} (\bar{y}) - g_{n,\tau_i} (\bar{y}) | \leq \sqrt{m} |\tau - \tau_i| \leq u,$$
for $\tau \in [\tau_i, \tau_{i+1}]$. Hence,
$$H_B (u, \mathcal{T}_n, L_2(P))  \leq  A_1 \log(u/\sqrt{n}),$$ for some constant $A_1 >0$ and for small $u>0$. As the class $\mathcal{G}_{n,k}$ is formed by product of the two classes  $\mathcal{S}_{n,k}$ and $\mathcal{T}_n$, the bracketing number for $\mathcal{G}_{n,k} $ is bounded above by,
$$ H_B (u, \mathcal{G}_{n,k}, L_2(P)) \leq A_0 u^{-1} + A_1 \log(u/\sqrt{n}) \leq A_2 u^{-1}.$$
In light of the above bound on the entropy, the first term in \eqref{c5eq:simpmn_et} can be shown to go to zero by arguing in the same manner as in the proof of Proposition \ref{c5prp:ratemu}.

For the second term in \eqref{c5eq:simpmn_et}, note that $|\Phi \left( {\sqrt{m} (\bar{Y} - \tau)} \right)  - \Phi \left( {\sqrt{m} (\bar{Y} - \tau_0)} \right)| \leq \sqrt{m} |\tau_0 - \tau|$. Hence,
\begin{eqnarray*}
\lefteqn{\left| ({M}_m (S,\tau) - {M}_m (S,{\tau_0}) ) -  ({M}_m (S_0,\tau) - {M}_m (S_0,{\tau_0}) ) \right| }\\
& =& \left| P_m \left[\left\{\Phi \left( {\sqrt{m} (\bar{Y} - \tau)} \right)  - \Phi \left( {\sqrt{m} (\bar{Y} - \tau_0)} \right)\right\}\left\{1_S(X) -1_{S_0}(X)\right\}\right]\right| \\
& \leq& \sqrt{m} |\tau_0 - \tau| \left| P_m \left|1_S(X) -1_{S_0}(X)\right|\right|.
\end{eqnarray*}
Thus, the second term in \eqref{c5eq:simpmn_et} is bounded by
$$2 (L_{\alpha} / \sqrt{n}) \sup_{S \in \mathcal{S}_{n,k}}   P_m |1_S(X) - 1_{S_0}(X)| \leq L_{\alpha} 2^{k+2} \delta_n / \sqrt{n} .$$
This is eventually smaller that $c(\gamma) 2^{k} \delta_n/3$ and hence, each term in the second sum of \eqref{c5eq:simpmn_et} is eventually zero. 
As $\alpha$ is arbitrary, we get the result.
\qed

\subsection{Proof of Proposition \ref{c5th:extnuntaureg}} \label{c5pf:extnuntaureg}
Note that $\sqrt{n}(\hat{\tau} - \tau_0) = O_P(1)$. So, given $\alpha > 0$, there exists $L_\alpha >0$ such that for $V_{n ,\alpha} = [\tau_0 - L_\alpha/\sqrt{n}, \tau_0 + L_\alpha/\sqrt{n}]$, $P[\hat{\tau} \in V_{n,\alpha}] > 1 - \alpha$. Let $\hat{S}_n (\tau)$ denote the estimate  of $S_0$ based on $\M_n(S, \tau)$. We have,
\begin{eqnarray*}
{ P^*\left[d(\hat{S}_n(\hat{\tau}), {S}_0) > \delta_n\right] }  & \leq & P^*\left[\delta_n < d(\hat{S}_n(\hat{\tau}), \hat{\tau} \in V_{n,\alpha}\right] + \alpha.
\end{eqnarray*}
 Following the arguments for the proof of Proposition \ref{c5prp:ratemureg}, 
the first term can be bounded by
\begin{eqnarray*}
\sum_{k\geq 0, 2^k \delta_n \leq 1 } P^*\left( \inf_{A \in \mathcal{S}_{n,k} } \M_n(A, \hat{\tau}) - \M_n({S}_0, \hat{\tau}) \leq 0, \hat{\tau} \in V_{n,\alpha} \right) 
\end{eqnarray*}
This is at most  
{\small \begin{equation} 
\sum_{k=0}^{k_n} P^*\left(\sup_{\substack{A \in \mathcal{S}_{n,k},\\ \tau  \in V_{n,\alpha}}} \left| (\M_n(S,\tau) - M(S)) -  (\M_n(S_0,\tau) - M(S_0)) \right| > \inf_{A \in \mathcal{S}_{n,k}} \left({M}(S) - {M}({S}_0) \right)  \right). 
\label{c5eq:shellmn_et_reg}
\end{equation}}
Note that $ M(S)  - M(S_0)  \geq c(\gamma) \lambda(S \triangle S_0)$ as earlier,
and hence \eqref{c5eq:shellmn_et_reg} is bounded by
{\small \begin{equation} 
\sum_{k=0}^{k_n} P^*\left(\sup_{\substack{A \in \mathcal{S}_{n,k},\\ \tau  \in V_{n,\alpha}}} \left| (\M_n(S,\tau) - M(S)) -  (\M_n(S_0,\tau) - M(S_0)) \right| > c(\gamma) 2^{k} \delta_n \right) .\nonumber
\end{equation}}
Moreover,
\begin{eqnarray*}
\lefteqn{\left| (\M_n(S,\tau) - M(S)) -  (\M_n(S_0,\tau) - M(S_0)) \right|} \hspace{1in}\\
&\leq& \left| (\M_n(S,\tau_0) - M(S)) -  (\M_n(S_0,\tau_0) - M(S_0)) \right| \\
& & + \left| (\M_n(S,\tau) - \M(S,\tau_0)) -  (\M_n(S_0,\tau) - \M_(S_0,\tau_0)) \right|.
\end{eqnarray*}
By the Lipschitz continuity of $\Phi$, we have 
\begin{eqnarray*}
\lefteqn{\left| (\M_n(S,\tau) - \M(S,\tau_0)) -  (\M_n(S_0,\tau) - \M_(S_0,\tau_0)) \right|}\hspace{1.5in}\\
& \leq & \sqrt{n h_n^2} |\tau - \tau_0| \left[\frac{1}{n} \sum_{k,l} \left|1_S(x_{kl}) - 1_{S_0} (x_{kl})\right|\right]\\
& \leq & \sqrt{n h_n^2} |\tau - \tau_0| \left[\lambda(S\triangle S_0) + O\left(\frac{1}{\sqrt{n}}\right)\right].
\end{eqnarray*}
Here, the last step follows from calculations similar to those in the proof of Lemma \ref{lm:constbound}. Consequently, for sufficiently large $n$,
\begin{eqnarray*}
\lefteqn{\sup_{\substack{A \in \mathcal{S}_{n,k},\\ \tau  \in V_{n,\alpha}}}\left| (\M_n(S,\tau) - \M(S,\tau_0)) -  (\M_n(S_0,\tau) - \M_(S_0,\tau_0)) \right|}\hspace{1.5in}\\
& \leq & (2 L_\alpha) h_n \left[2^{k+1} \delta_n + O\left(\frac{1}{\sqrt{n}}\right)\right] < \frac{c(\gamma)}{2} 2^{k}\delta_n. 
\end{eqnarray*}
Hence,
\begin{eqnarray*}
\lefteqn{\sum_{k=0}^{k_n} P^*\left(\sup_{\substack{A \in \mathcal{S}_{n,k},\\ \tau  \in V_{n,\alpha}}} \left| (\M_n(S,\tau) - M(S)) -  (\M_n(S_0,\tau) - M(S_0)) \right| > c(\gamma) 2^{k} \delta_n \right) }\\
&\leq& \sum_{k=0}^{k_n} P^*\left(\sup_{\substack{A \in \mathcal{S}_{n,k},\\ \tau  \in V_{n,\alpha}}} \left| (\M_n(S,\tau_0) - M(S)) -  (\M_n(S_0,\tau_0) - M(S_0)) \right| > \frac{c(\gamma)}{2} 2^{k} \delta_n \right). 
\end{eqnarray*}
The above is a probability inequality based on the criterion with known $\tau_0$. This can be shown to go to zero by calculations identical to those in the proof of Proposition \ref{c5prp:ratemureg}. As $\alpha>0$ is arbitrary, we get the result.
\qed

\bibliography{References}

\end{document}